# Bubble-assisted micromixing via thermally excited intrinsic air within microfluidic systems

**Authors' names:**

Seyed Shahriar Riazi[a]

Seyed Mostafa Hosseinalipour[a]

[a] Energy, Water, and Environment Research Center, School of Mechanical Engineering, Iran University of Science and Technology, 16846-13114 Tehran, Iran. ss7riazi@gmail.com, alipour@iust.ac.ir.

**Corresponding author:** Seyed Mostafa Hosseinalipour (alipour@iust.ac.ir).

## Abstract:

The micromixing process in microfluidic devices is of central importance in quite a few applications, with microfluidic studies on carbon capture and environmental monitoring being two examples. High surface area to volume ratio in microscale flows outweighs the role of interfacial tension manipulation by means of introducing a secondary phase to the main flow in order to augment the typically diffusion-dominant mass transfer operation. In this paper, we introduce an easily integrable continuous flow micromixing scheme implemented on a simple Y-type microchannel based on thermally excited intrinsic air within microchannel. Thanks to the direct contact between the sputtered thin platinum microheater and the co-flowing streams, trapped air in the liquid and/or tiny crevices is employed to generate elongated bubbles with a millisecond lifespan, consuming less than 0.4 W of power and thereby enabling homogeneous mixing. The performance of the micromixer is characterized in terms of the mixing index (MI), and it is shown that immediately ahead of the microheater, the MI exceeds 95% for side-by-side flowing of water at different overall flowrates, namely 4, 10, and 20 μL/min. Interpreting the experimental results, supplemented by scaling arguments, we quantitatively describe the ephemeral role of emerging elongated bubbles in the micromixing process in which the Marangoni as well as Weber numbers are recognized as the characteristic local non-dimensional groups while the enhancing role of tiny daughter microbubbles on the micromixing downstream of the microchannel can be reflected in the Péclet and the Capillary numbers.

# 1 Introduction

Miniaturization of fluidic devices and flow of fluids in diminutive channels—where the characteristic lengths go below the thickness of a human's hair, which is called microfluidics (MFs)—have been revolutionized analytical techniques as well as many industrial processes [1]. Minimal consumption of reagents, automation, integration, and repeatability make the use of MFs popular in a myriad of chemical and biochemical applications, which themselves consist of many unit operations. Mixing, of the most frequent unit operations, is of critical importance in micro-scale applications, including microreactors, micro-extraction of the precious metals as well as rare-earth elements in microgravity conditions, carbon capture, pollutant monitoring, and multimaterial printing [2–11].

Despite the many beneficial features of MFs, the design of micromixers faces two main challenges: the intrinsic tendency of low-Reynolds number flows to be reversible, and the high Péclet number [1,12]. Micrometer scales and the smallness of fluid velocities suppress the likelihood of turbulence. Furthermore, Taylor-Aris dispersion tends to happen (owing to the parabolic velocity profile of laminar flows), meaning that not only axial mixing occurs and mixing length goes beyond several centimeters, but also it detrimentally lessens the resolution of analytical investigations [1,12]. Hence, it is necessary to increase the disruption of liquid-liquid interfaces and impose stretching and folding to them to intensify the local and slow process of diffusion [13]. In this regard, various strategies have been considered which can be classified into two main categories: passive and active [14,15]. In the former case, mixing is achieved through adapting the microchannel geometry and/or imposing centripetal effects on the streams. Nonetheless, several difficulties such as enormous pressure drop, weak controllability over the quality of mixing, and complexities in microfabrication techniques should also be addressed in their design [12,16].

On the other hand, active micromixers benefit from an external source for altering the velocity field and intensifying recirculatory motions. High surface area to volume ratio—one of the key features of MFs—makes the physics of flow in microchannels entirely different from routine large-scale systems, where the modulation of interfacial energy between liquid-gas interfaces has been proved to be a robust way for many purposes in MF devices [17,18]. Surface tension at a liquid-gas interface is a function of temperature and chemical composition in which both can be utilized to induce a gradient of surface tension disturbing the mixing barriers. Thermal-Marangoni or thermocapillary are the names allocated to the stresses that stem from a gradient of temperature [12,19].

In this regard, Darhuber et al. [20] designed an open MF device based on indirect heating by microheaters mounted below the substrate, and gradual mixing occurred due to the gradient of surface stresses at the liquid-air interface. Although one of the first micromixers took advantage of Marangoni effects, vulnerability to contamination and evaporation as well as slow response time faced its applicability with challenges. In another research, by inducing intense laser beams by means of an expensive, standalone apparatus on a small area of laminar co-current flows, Hellman et al. [21] proposed an ultrafast micromixing scheme. Such a transient, extreme heat flux instantaneously drives the fluids into a non-equilibrium condition, creating unstable bubbles. Despite its susceptibility to failure, this research highlights the potential of bubbles to bring about segmented flow patterns for micro-scale transport phenomena [22,23].

It is noticeable to allude to some viewpoints that consider bubbles—omnipresent on microchannel surfaces—as "the villain of microfluidics" [24–26]. The supersaturation condition of aqueous streams in MF applications and trapping of Harvey nuclei [27] on surface nanoscale cracks are supposed to be the reasons behind the high probability of bubble nucleation within microchannels [25]. Furthermore, the widespread use of intrinsically hydrophobic, high gas permeable polymers such as polydimethylsiloxane (PDMS) in the fabrication of MF devices tends to entail the wettability of surfaces in the Cassie-Baxter state for trapped air pockets [25,28]. Thus, although such a condition might seem detrimental to the controllability of MF devices at first glance, the systematic utilization of these endogenous air pockets is an opportunity for designers to augment micromixing as well.

In addition to the previous research, there is a penchant for taking advantage of multiphase systems by introducing bubbles from an external source to the main streams [16,29,30]. Günther et al. [16] designed a gas bubble-assisted MF device for mixing; an inert gas was introduced to the streams, and by regulating the flow rates, discrete segments of liquids and bubbles were formed. They visualized velocity fields and disclosed vortical flows induced by bubbles. However, due to the chemical difference between the main flows and the introduced bubbles, addressing a mechanism for separating liquids and the inert gas at the end of the process seems necessary. In this regard, it has been shown that an extra separation unit is needed [16,29,30].

In this paper, a micromixing scheme based on thermally excited intrinsic bubbles within microchannel crevices as well as flowing liquids is represented for a simple Y-type square microchannel made of PDMS (as shown in Fig. 1) that obtains thorough mixing by consuming less than 0.4 W of power and can be compatibly integrated with a wide range of prefabricated MF masks. By virtue of direct contact between microheaters and streams, trapped air in liquids and/or tiny cavities of the microheaters substrate are employed to create the segmented flow, and hence, neither external inert gas sources nor a separation unit is required (compared to some of the afore discussed micromixers). The effects of various aqueous solutions with different viscosities and different flow rates on the mixing quality have been experimentally studied, and the results are thoroughly interpreted and supplemented with physical scaling arguments. Moreover, due to the accessibility and portability of the proposed micromixer, it is expected that the scheme facilitates micromixing challenges for chemical analysts.

## 2 Experiment and methodology

### 2.1 Materials

In procurement and preparation of the materials, considerable attention was devoted to satisfying the analytical reagent grade. A variety of aqueous solutions, ranging from deionized water to different weight fractions of 5 %, 10 %, 15 %, and 20 % of glycerol, was deployed. Glycerol was purchased from Dr. Mojallali Industrial Chemical Complex, Iran. Methylene blue (MB) dry dye (acquired from Merck, Germany) was fed into one of the streams so as to have distinguishable streams. PDMS Sylgard 184 kits were obtained from Dow Corning, USA. Photoresist (SU-8 2050), developer, and positive photoresist (Shipley S1813) were supplied by MicroChem, USA.

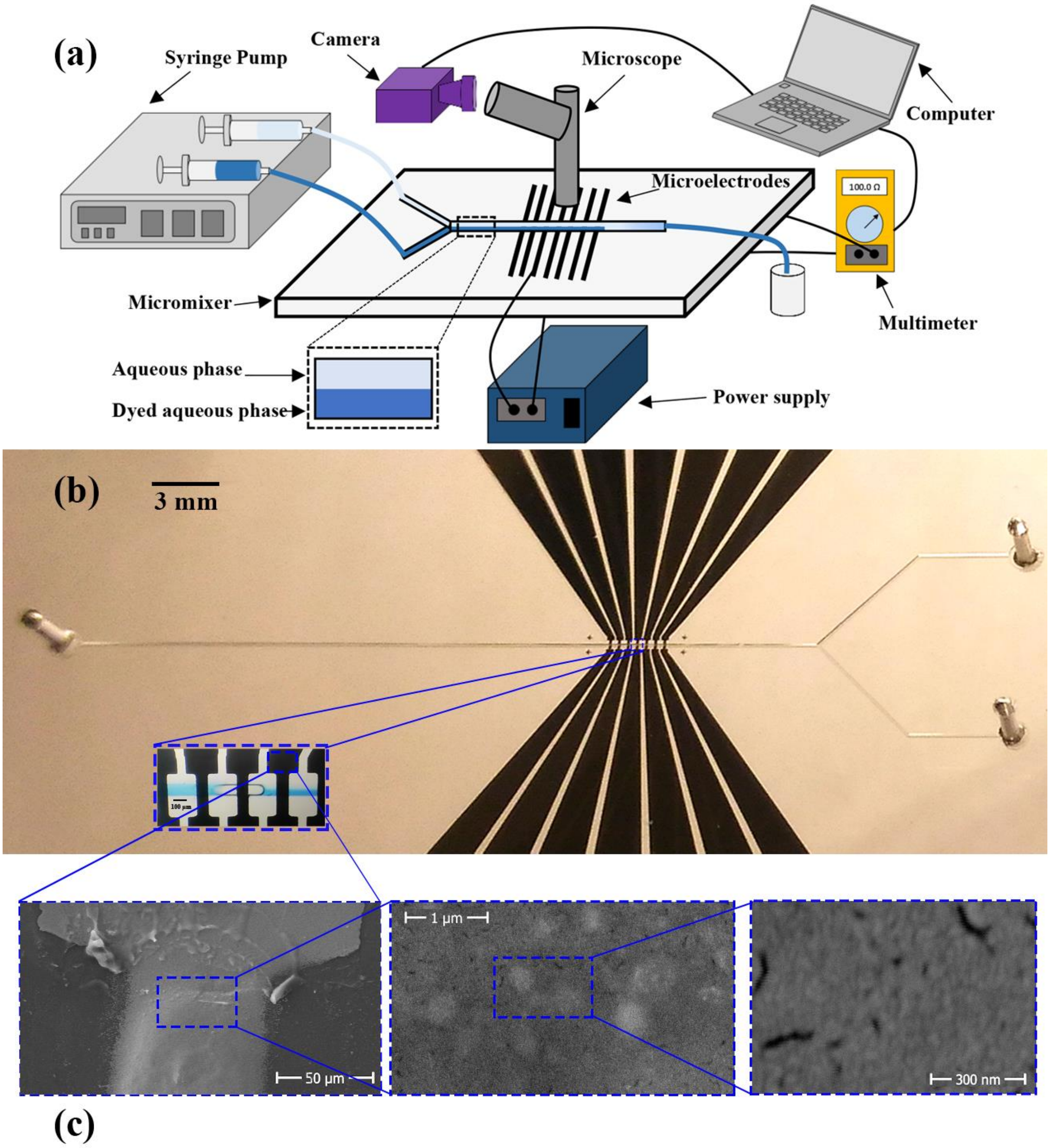

**Fig. 1.** Micromixing scheme and the experimental setup: (a) schematic diagram; (b) fabricated micromixer; and (c) FE-SEM images of the surface of microelectrodes.

## 2.2 Microchannel fabrication

The micromixer consists of a microchannel made by PDMS and a glass substrate, in which platinum microelectrodes (μEs) are sputtered. The microchannel has a square cross-section of side 100 μm and is fabricated using standard soft lithography techniques; a silicon wafer of a thickness of 500 μm and 7.62 cm in diameter was spin-coated with SU-8 at 1600 rpm for 30 s, and then, baked at 95 °C for 2 min to obtain the mold. Then, it was exposed to 1.4 mW.cm$^{-2}$ ultraviolet rays for 120 s to print the photomask pattern of the microchannel and then baked at 95 °C for 10 min. PDMS prepolymer was mixed with a curing agent at a 10:1 weight ratio in order that it can be used for pouring onto the mold. Before pouring and by means of a vacuum desiccator, air bubbles are removed from the mixture. Afterward, the mixture was heated on a hotplate at 90 °C for 8 min. After lifting off the microchannel from the mold, the inlets and outlet were punched by a biopsy punch.

## 2.3 Microelectrodes fabrication

Seven μEs are considered for direct contact heating of the streams as well as temperature measurement, as shown in Fig. 1 (b). The width of μEs is 90 μm, and their center-to-center distance is 400 μm. The μEs were extended to the exterior part of the microchannel (approximately 1 cm distance from the microchannel) so that they would easily connect to the power supply system. After designing a positive photomask for the array of μEs, S1813 was spin-coated onto a glass substrate at 3000 rpm for 30 s to obtain a 1 μm layer of the photoresist. The glass was baked at 95 °C for 2 min and then transferred into a mask aligner apparatus, in which it was exposed to 1.4 mW.cm$^{-2}$ ultraviolet rays for 60 s to print the photomask pattern of the μEs. By utilizing the sputter deposition technique, a primary titanium layer of 30 nm as well as a final platinum layer of 100 nm was sputtered on the glass; the sputter deposition technique was done by placing targets (metals) into the rotation magnet sputtering apparatus under high vacuum ($1.9\times10^{-4}$ mbar) condition. By inserting argon gas into the chamber and increasing the pressure to $8\times10^{-3}$ mbar, atoms from the targets were removed and coated on the pattern. Sputtering of titanium and platinum is done by consuming 60 and 38 W for 5 and 7 min, respectively.

The substrate was baked at 450 °C for 1 h under a furnace with vacuum condition (0.4 mbar), thereby reducing the resistance of μEs. Finally, the microchannel and the glass were bonded together under a plasma bonding chamber and exposed to 40 W oxygen plasma at 0.8 mbar for 1 min. In order to straighten the bonding, the device is also heated on a hotplate at 90 °C for 5 min. Following the completion of the experiments, Field Emission Scanning Electron Microscopy (FE-SEM) images were taken to observe qualitatively the morphology of the surface of μEs, as shown in Fig. 1 (c). In this regard, uncertainty in the height of the microchannel as well as the μEs' width are estimated to be below 5 μm and 2 μm, respectively.

## 2.4 Experimental setup and procedures

An experimental setup, as shown in Fig. 1 (a), is set up to investigate the mixing characteristics of the MF chip. The MF chip consists of platinum µEs; the third µE is used for heating—supplied by a low-voltage power supply system (Dazheng PS-305D)—and three µEs, one behind it (the second one) and two ahead of it (the fifth and sixth µEs), are responsible for the temperature measurement. There also exist other µEs considered for future studies. The µEs have direct contact with the side-by-side flow of two streams, which were pumped via a syringe pump (Samane Tajhiz Danesh, Iran) through a Y-shaped microchannel. The syringe pump has an injection accuracy of 0.1 µL/min. Mixing performance was monitored using a microscope, Olympus CX21, equipped with 240 fps video camera, Xiaomi Redmi Note 9 Pro.

The inlet aqueous solutions were pumped through the microchannel at an approximate ambient temperature of 20 °C. Between every two consecutive experiments, the power supply system is turned off, and the syringe pump continues to run for at least 10 min to cool down the microchannel into the ambient temperature and remove bubbles adhering to the microchannel walls. In each test, inserted voltage increases from zero by half a volt in a gradual manner (at least 2 min interval between successive steps) until the bubble emergence. In order to gain an estimation of the temperature distribution in the proximity of the nucleation site, the resistances of utilized µEs were recorded via multimeters.

**Table 1**

Physical properties of pure water and glycerol in water solutions (GSs).

| Solution | Pure Water | | 5 wt% GS | | 10 wt% GS | | 15 wt% GS | | 20 wt% GS | |
|---|---|---|---|---|---|---|---|---|---|---|
| | 20 °C | 35 °C | 20 °C | 35 °C | 20 °C | 35 °C | 20 °C | 35 °C | 20 °C | 35 °C |
| Density (ρ) [kg/m$^3$] | 998 | 992 | 1009 | 1005 | 1022 | 1017 | 1035 | 1030 | 1048 | 1043 |
| Dynamic Viscosity (µ) [mPa.s] | 0.97 | 0.72 | 1.13 | 0.82 | 1.31 | 0.93 | 1.52 | 1.05 | 1.68 | 1.23 |
| Surface Tension (σ) [mN/m] | 72.4 | 70.3 | 72.2 | 70.1 | 71.6 | 69.6 | 71.3 | 69.2 | 71.0 | 68.9 |

## 2.5 Physical properties

The physical properties of used aqueous solutions—including density (ρ), dynamic viscosity (µ), and surface tension (σ)—are listed in Table 1. The experiments were conducted at two different temperatures, specifically 20 and 35 °C, during which the physical properties exhibited negligible variations in line with the literature [31,32]. Surface tensions as well as densities are measured by KSV Sigma 700/701 tensiometer. Since there is an almost linear relation between surface tension and temperature over a wide temperature range for water and the other aqueous solutions [33], the temperature-coefficient of surface tension ($\sigma_T$) is calculated based on the values of σ at two different temperatures. It is estimated to be -0.14 mN/m°C for all solutions. Dynamic viscosities are measured by ViscoStar viscometer. Since solved MB in aqueous solutions are at infinite dilution, the diffusion coefficient ($D_{AB}$) of solute MB in water was estimated $3.8\times10^{-10}$ $m^2/s$, according to Ref. [34]. Additionally, $D_{AB}$ for glycerol solutions was adapted using the Wilke-Chang equation, incorporating viscosity changes and a simple mixing rule to modify the association parameter and molecular weight [35].

## 2.6 Temperature-resistance calibration

Since there is an appropriate linear relationship between temperature and resistance for Platinum over a wide temperature range, the µEs can also be recruited as thermistors to estimate the wall temperature. Due to the special fabrication method of the µEs, each of them must be calibrated separately. In order to find the thermal-resistance calibration of the µEs, the IKA RH BASIC 2 hot plate was used to control the temperature from 20°C to 90°C in 10°C increments. The resistance was recorded at each temperature step, and each test was extended for 10 min to ensure steady-state conditions before data collection. The calibration for used µEs is presented in Table 2.

**Table 2**

Temperature versus resistance calibration. The microelectrodes are numbered from left to right according to Fig. 1.

| Microelectrode | T (°C) vs. R (Ω) Calibration | Coefficient of Determination |
|---|---|---|
| 3 | $T = 1.7967\mathrm{R} - 272.27$ | 0.9840 |
| 5 | $T = 2.1010\mathrm{R} - 315.03$ | 0.9946 |
| 6 | $T = 2.1327R - 359.10$ | 0.9920 |

## 2.7 Quantification of mixing quality

The quality of the mixing is determined through the mixing index ($MI$):

$$MI = 1 - \frac{\sqrt{\left(\frac{1}{n}\right)\sum(I_i - \bar{I})^2}}{\bar{I}}, \quad (1)$$

where $I_i$ is the color intensity of each pixel in the HSI (Hue, Saturation, Intensity) color model, $\bar{I}$ is the average color intensity, and $n$ denotes the total number of cells. After rearrangements and extracting images from the recorded videos by Adobe Photoshop, a domestic image analysis code written in MATLAB processes the images and calculates mixing indices.

## 2.8 Uncertainty analysis

Throughout this paper, we have made every effort to remain consistent with the procedures of [36] in representing the experimental results and their associated uncertainty. In the calculations of non-dimensional numbers (Sections 3.5 and 3.6), uncertainty propagation is estimated using the following formulae for a functional relation of $f = f(x_1, x_2, \dots, x_n)$ where $f$ is the function and the $x_i$ are the variables:

$$\delta f = \sqrt{\left[\left(\frac{\partial f}{\partial x_1}\right)\delta x_1\right]^2 + \left[\left(\frac{\partial f}{\partial x_2}\right)\delta x_2\right]^2 + \cdots + \left[\left(\frac{\partial f}{\partial x_n}\right)\delta x_n\right]^2}. \quad (2)$$

In Eq. (2), $\delta f$ and $\delta x_i$ refer to the uncertainties in the corresponding variables, whereas $\frac{\partial f}{\partial x_i}$ denotes the partial derivative based on the functional relation.

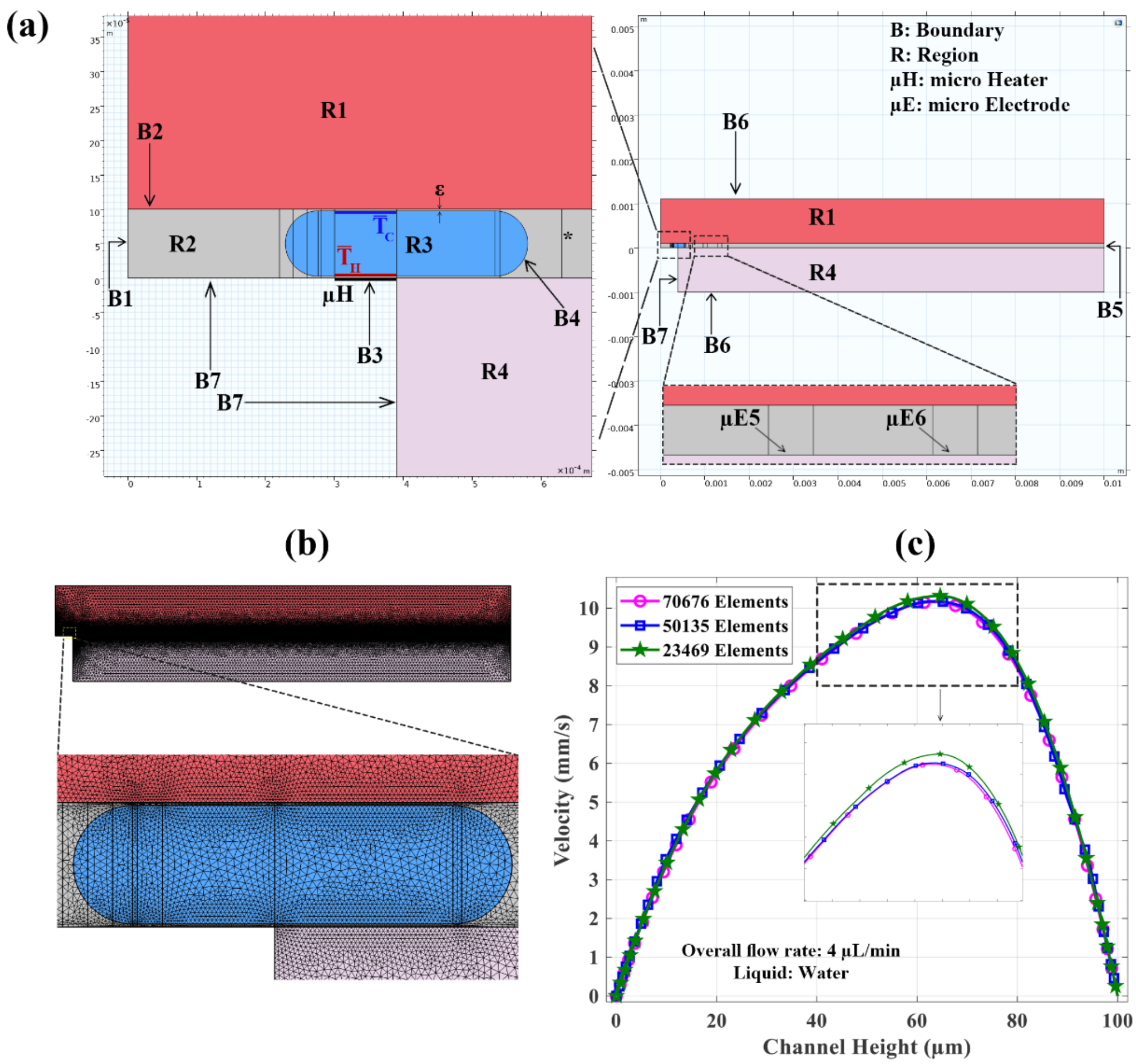

**Fig. 2.** (a) Computational Domain and specification of boundary conditions. The line average temperatures, $\overline{\boldsymbol{T}}_{\boldsymbol{C}}$ and $\overline{\boldsymbol{T}}_{\boldsymbol{H}}$, are used in the Marangoni number. (b) Representation of the mesh used in the simulations. (c) The grid independence study based on the velocity profile versus the channel height ahead of the bubble.

## 2.9 Numerical simulation

To supplement the thermometry study carried out using µEs and to find out the details of temperature distribution in the vicinity of emerged bubbles, a two-dimensional steady-state model based on the finite elements method were developed by using the COMSOL Multiphysics 5.6 software with the prescribed boundary conditions shown in Fig. 2. The simulations were performed for various solutions at two overall flow rates of 4 and 10 µL/min. The particular interest behind performing this numerical simulation is to obtain an acceptable approximation for the temperature difference between the microheater temperature, $\overline{T}_H$, and the colder temperature, $\overline{T}_C$, (as shown in Fig. 2 (a)). The simulation results are mainly useful in the calculations related to the Marangoni number (see Section 3.5.1). For this purpose, we used experimental data, including the length of the bubbles, for simulation. It is important to note that we do not attempt to fully simulate all of the complex physics involved in our thermally induced bubble-assisted micromixing strategy. Instead, we use experimentally derived geometries and boundary conditions to extend our thermometry capabilities to regions where direct measurements are experimentally inaccessible.

The dimensions of the microchannel and the heater are considered similar to those in the experimental setup. For the PDMS and glass regions, their heights are chosen such that the depth of heat diffusion can reach its maximum while avoiding an excessively large computational domain. Material properties for PDMS and glass are taken from COMSOL's glossary, while those for aqueous solutions are manually input based on Table 1 or, when unavailable there, from relevant references [31,32].

In Fig. 2 (a), R1, R2, R3, and R4 represent the regions of PDMS, aqueous solutions, the elongated bubble, and the glass substrate, respectively. The nonisothermal single-phase incompressible laminar flow was utilized for aqueous solutions, and the boundaries denoted by B's in Fig. 2 (a) were set as follows: (B1) The inlet was set to be a fully developed flow with determined average velocity. (B2) At the boundary between the liquid flow and solids, the no-slip boundary condition was supposed. (B3) The constant heat flux was assigned. (B4) Along the liquid-air interface, a stationary free surface was adjusted. (B5) The pressure outlet boundary condition was set. (B6) The external solid boundaries were assumed to be maintained at a constant ambient temperature of 20 °C, except for B7, where it was assumed to be thermally insulated. By means of a user-controlled mesh (more than 95% triangular), the computational domain was meshed, as shown in Fig. 2 (b). The grid independence study based on the velocity profiles versus the channel height at the cross-section indicated by a star in Fig. 2 (a) was performed for three different numbers of elements as shown in Fig. 2 (c).

# 3 Results and discussion

## 3.1 Micromixing regimes

The low-Reynolds number flow regime in microchannels and the axial mixing hinder homogeneous mixing in the case of adjacent streams, as can be seen in Fig. 3 for cold cases (i.e., zero applied voltage). When the microheater is off, the aqueous stream reaches the end of the microchannel, while the mixing index does not exceed 75 %. Briefly comparing the MIs when water flows with overall flow rate of 4 μL/min, one can find out that an increase of 15 % in the MI at the end of microchannel is due to the dispersion occurring after the coflowing continues for extra 4 mm. After turning the microheater on, it is observed that below a critical voltage the mixing enhances gradually. The mixing enhancement in this primary regime—the blue-highlighted region in Fig. 3, which is labeled as the Heat-Induced Mixing Regime (HIMR)—is mainly due to the temperature rise of coflowing streams and its subsequent effect on physical properties, compared to cold cases. The dashed lines in Fig. 3 denotes to the MIs ahead of the microheater while the solid lines refer to the MIs at the end of microchannel.

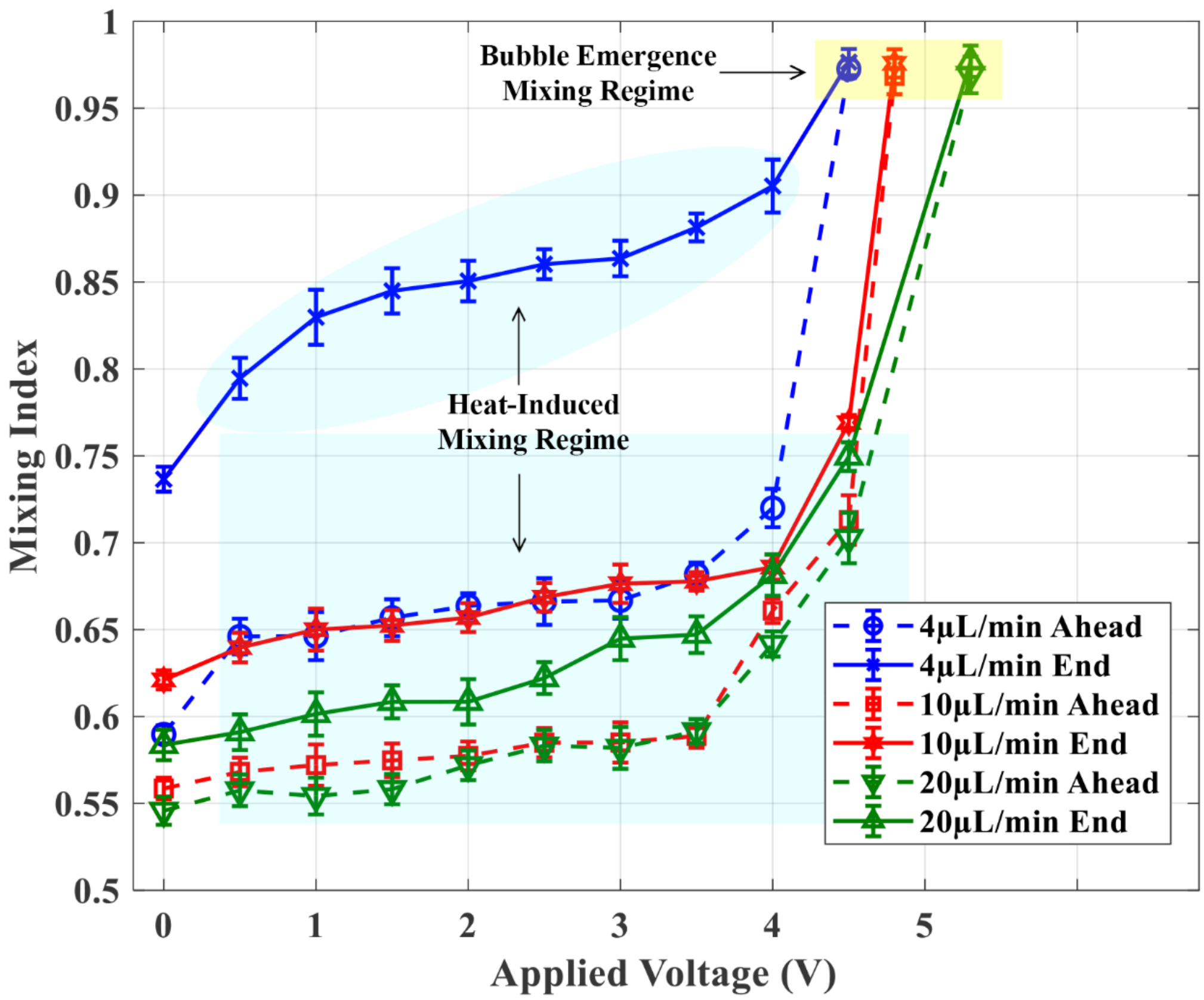

**Fig. 3.** Mixing index (MI) versus applied voltage for water ahead of the microheater and at the end of microchannel for different flow rates.

The mixing enhancement achieved through the HIMR is still far from a thorough mixing, and similar to the cold cases, the MI at the end of microchannel is greater than its counterpart ahead of the microheater. Furthermore, our investigation reveals that depending on the flowing liquid, there is a critical voltage above which elongated bubbles emerge. This regime, characterized by the Bubble Emergence Mixing Regime (BEMR) and shown by the yellow-highlighted region in Fig. 3, introduces a fast, thorough micromixing mechanism (see also the supplementary video). More specifically, immediately ahead of the microheater, the MI exceeds 95 % for side-by-side flowing of water for all flow rates, while consuming less than 0.4 W of power. Our use of the phrase "ahead of the microheater" refers to a region located approximately 200 to 700 μm downstream from the microheater, depending on the final observed length of the bubbles and the visibility allowed by the proximity of the microelectrodes, which appear black in the images. In Fig. 4 (a), time intervals at the onset of bubble emergence at the critical voltage are shown, where the third μE works as the microheater while the fifth and sixth μEs are used for thermometry purposes. Considering examined flow rates and various aqueous solutions, we observed that the critical voltage is within the range of 4.3-7.8 V with the subsequent electric current 0.02-0.04 A.

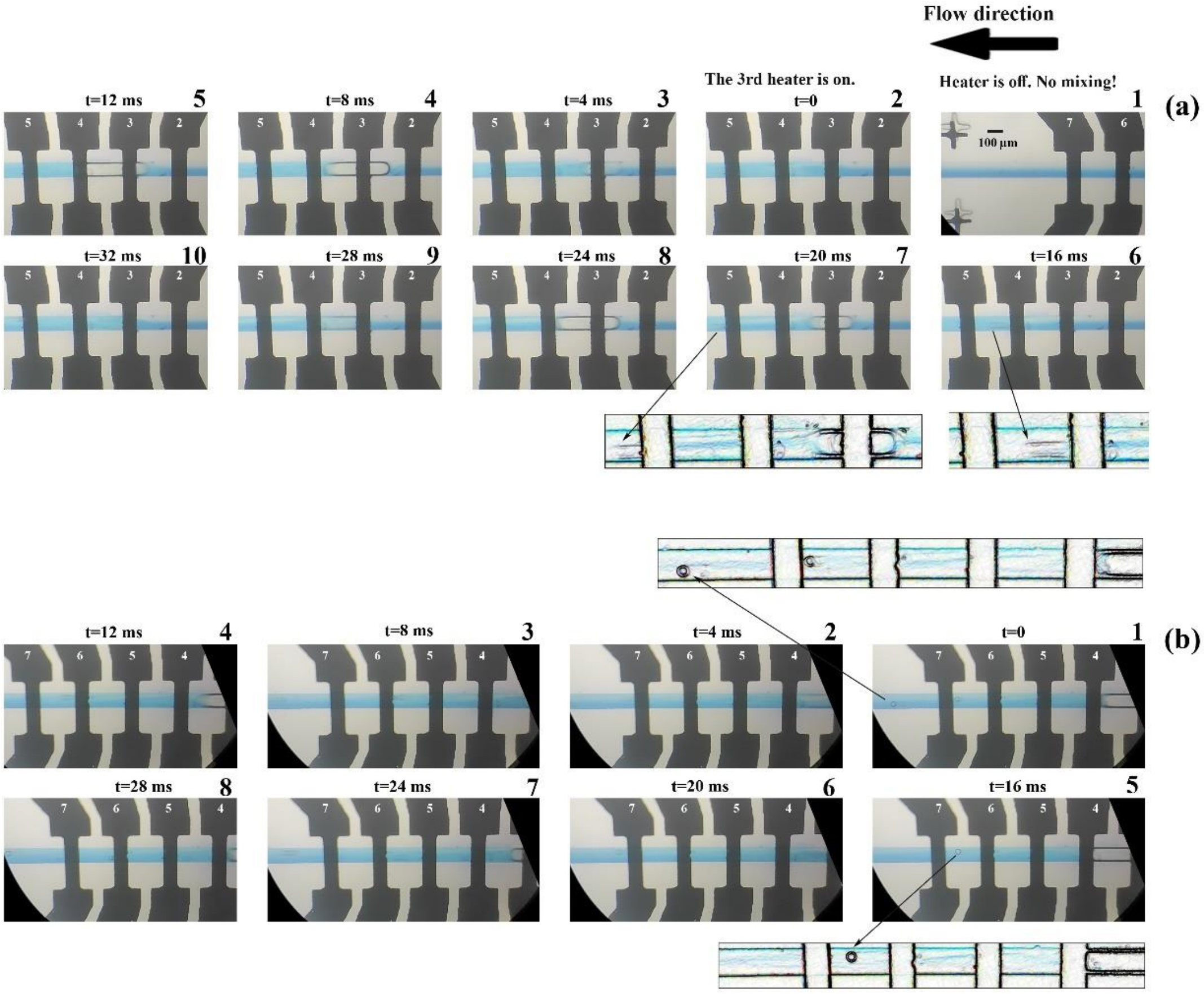

**Fig. 4.** Top view of the contiguous flow of water with an overall flow rate of 10 μL/min: (a) gradual emergence of the elongated bubble on the microheater surface; and (b) its pinch-off to tiny spherical daughter bubbles moving downstream. Edge detection is applied to characterize the daughter bubbles formation with improved clarity.

The details of critical voltages for various aqueous solutions as well as different flow rates are presented in Fig. 5. Accordingly, when the viscosity of aqueous solutions increases (by increasing the weight percent of glycerol), the critical voltage for obtaining the BEMR is increased to an extent that it is at least 1.5 times greater for the 20 wt% solution compared to that of pure water for a specific flow rate, while the MI in both situations (ahead of the microheater and at the end of microchannel) slightly decrease.

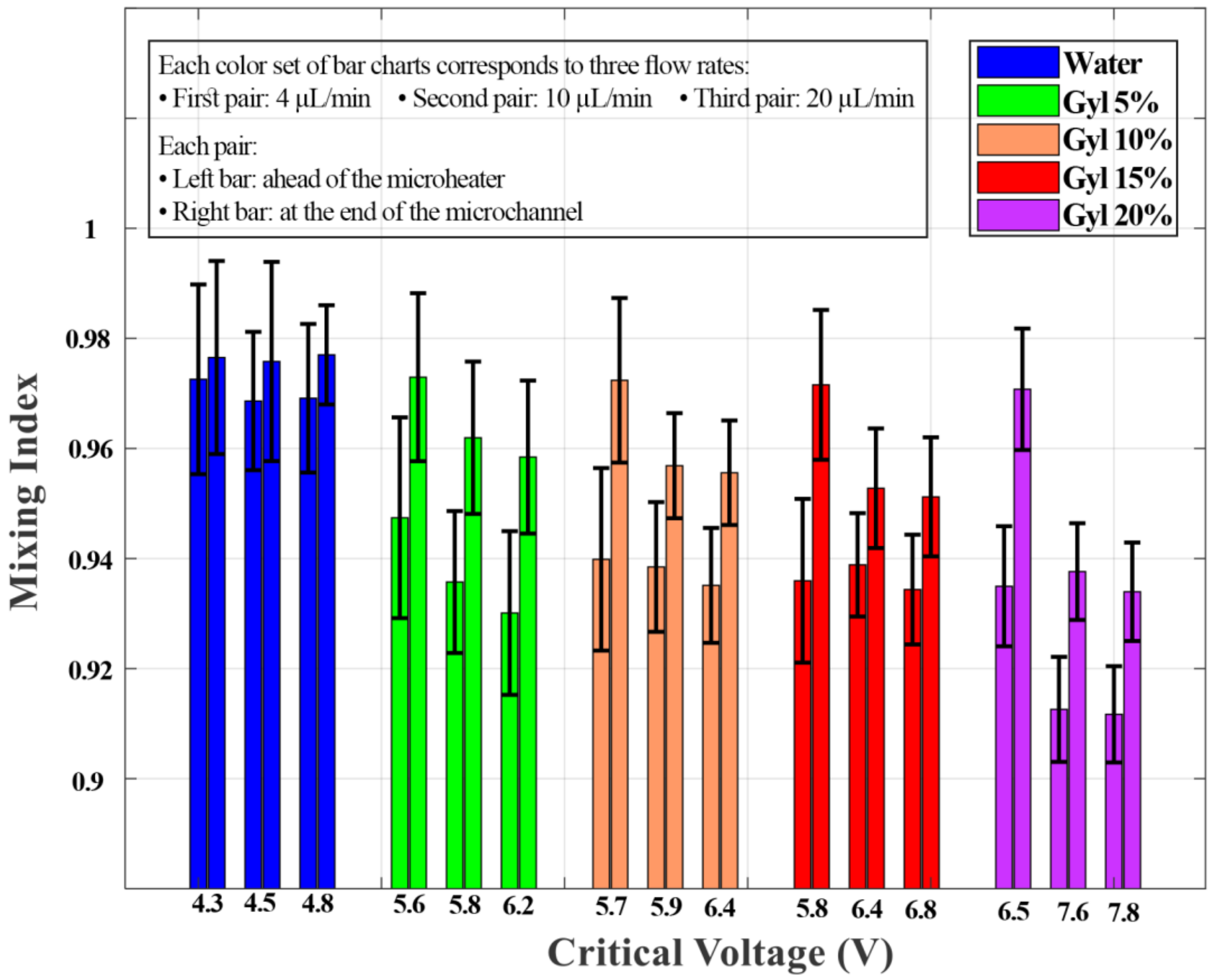

**Fig. 5.** Mixing index (MI) versus critical voltage for various flow rates of aqueous solutions. In each pair, the left and the right bars show the MIs ahead of the microheater and at the end of the microchannel, respectively.

As depicted through an edge detection filter in the sixth and seventh micrographs of Fig. 4 (a), the pinch-off and the microjetting process can be considered as a substantial mechanism in disrupting the liquid-liquid interfaces; the elongated bubble reaches a condition that it has to disintegrate into a microjet and satellite daughter spherical bubbles where they act as a means for stirring the otherwise diffusion-dominated streams. We believe that the presented micromixing strategy can be simply adopted for the prefabricated MF masks where it can robustly provide the beneficial effects of segmented flow regimes. Therefore, to better understand the intricate phenomena associated with the BEMR, an understanding of the temperature field around the elongated bubble as well as the governing forces during its lifespan is critically important, and thus, further discussion on this topic is presented in the following sections.

## 3.2 Description of hydrodynamic flow patterns

The emergence of the bubble drives the flowing streams to squeeze through the vanishing thickness, lubrication film between microchannel wall and the elongating bubble during its lifespan, similar to the classic Bretherton bubble problem [19,37–40]. Separation and recombination of liquid passages are also effective in the micromixing process. According to specific thermo-hydrodynamic property of the flowing streams, the elongation reaches a maximum value with a subsequent microjetting and breakup to tiny spherical daughter bubbles moving downstream, as shown in Fig. 4 (a) and (b) by means of edge detection filter.

The formation of the elongated bubbles not only enforces a curvature to the flat horizontally aligned liquid-liquid interfaces and disrupts them but also acts as a source of thermal-Marangoni stresses at the bubble-liquid interfaces. In other words, since the emerged bubbles in the present study inherently stem from the thermal manipulation of the flowing liquids, the interfacial temperature gradient across and along the elongated bubbles highlights the role of thermal-Marangoni stresses competing with other well-known forces, namely the momentum force of the growing bubble, the viscous force and the capillary force during the bubble emergence, elongation, and pinch-off cycle. It was theoretically proven by Brøns et al. [41] that the existence of an interfacial gradient for either the surface tension or the bubble-liquid curvature can act as active vorticity generator which thereby enhancing micromixing. In this regard, the effect of sole curvature variation was experimentally shown by Günther et al. [16] through trapping liquid segments between bubble slugs in their micromixing investigation.

We detected the bubble emergence, elongation, and final pinch-off cycle specifically for two overall flow rates of 4 and 10 μL/min in our investigation, where the elongated bubble during its millisecond-long lifespan does not move downstream with the background flow. The details of measured critical voltage ($CV$), lengths ($L$), growth rates ($\dot{L}$), and nucleation period ($\tau$) of elongated bubbles in these two overall flow rates for water and various glycerol-water solutions are summarized in Table 3. As the critical voltage for nucleating bubbles has not increased significantly (i.e., for water as well as 5 wt% and 10 wt% glycerol solutions) and has remained in the range of $4.3 - 5.8\,V$, the nucleation period is mostly controlled by two important variables, namely the overall flow rate and the weight composition. For example, in the case of water, the nucleation period decreases approximately by half from $32 \pm 4\,ms$ as the overall flow rate increases from 4 μL/min to 10 μL/min which is primarily due to the intensification of upstream convective effects. Furthermore, it should also be noted that in this scenario, the introduction of glycerol to the aqueous solution also has a slight delaying effect on nucleation of bubbles as long as the critical voltage does not surpass the $5.8\ V$. We attributed the latter effect mostly to the increase of viscosity of 5 wt% and 10 wt% glycerol solutions with respect to water by 17 % and 35 %, respectively.

However, an increase in the content of glycerol to higher levels corresponds to higher levels of critical voltage—for example, a 35 % and 69 % increase for 20 wt% GS compared to water at 4 and 10 μL/min, respectively. As discussed by Volk et al. [26], glycerol inherently has a smaller capacity for dissolving air, a fact that stems from its typically lower vapor pressure values. Therefore, increasing the glycerol content makes the aqueous solution deficient in dissolving air and demands higher amounts of input energy for bubble nucleation. After the occurrence of bubble nucleation, the constantly applied higher critical voltages decrease the nucleation period slightly. This effect is somewhat analogous to the reduction in bubble formation time period observed when increasing the flow rate of the dispersed fluid in T-junction microfluidic bubble generation [42], where increased energy or momentum input similarly accelerates the detachment process.

According to Table 3, both $L$ and $\dot{L}$ increase with the flow rate and the glycerol-to-water ratio. Before delving into the reasons behind the increasing relationship between $L$ and the glycerol-to-water ratio, we review the increasing trend between $\dot{L}$ and the glycerol content. As discussed in previous paragraphs, the applied heat which is necessary for the bubble nucleation increases with the glycerol content. After the occurrence of nucleation, this applied heat is consumed in every cycle of bubble generation partly to shed encapsulated air and partly to evaporate the local flowing fluid. Whichever mechanism is more dominant, one can conclude that the applied heat from the microheater, $\dot{Q}_{\mu H}$, is linearly proportional to the mass flow rate of generated bubble, $\dot{m}_{GB}$, which can be scaled as $\rho_B h^2 \dot{L}$ ($\rho_B$ is the density of bubble and $h$ is the height of microchannel). Consequently, the more applied voltage corresponds to a higher growth rate of the generated bubbles.

A continuation of the analogy with the T-junction microfluidic bubble generation further allows for the rationalization of the observed variations for $L$ in Table 3. As discussed by Garstecki et al. [42,43], in the squeezing regime of bubble generation in T-junction microfluidics, the Capillary number, $Ca = \mu U/\sigma$, must be of the order $O(10^{-2})$ or below where μ and U are dynamics viscosity and the average velocity of the background flow, respectively. As demonstrated in Section 3.6, the Capillary number in our study falls within this range. After comprehensive discussion by Garstecki et al., they outweighed the role of the ratio of flow rates of the dispersed and the continuous (carrier) fluids over shear stresses in predicting the length of generated bubbles. They found the length of the generated bubble relative to the height/width of the microchannel is almost linearly dependent on the ratio of flow rates of the dispersed fluid (in our analogy, the generated bubbles) and the continuous fluid (in our analogy, the background aqueous solution). Utilizing this relationship for the present study, we can deduce that $L/h = 1 + \beta\, \dot{Q}_{\mu H}/Q_{BG}$, where $Q_{BG}$ is the flow rate of the background aqueous solution and $\beta$ is a dimensional scaling constant. Consequently, the more applied voltage corresponds to a greater length of the generated bubbles.

On the other hand, when it comes to higher overall flow rates, namely 20 μL/min in our study, the flow hydrodynamics change to the slug flow pattern, as shown in Fig. 6, and the pinch-off as well as the microjetting process is shifted somewhere downstream. The isothermal slug flow pattern was investigated thoroughly [16,17,29,30,44] and it has been proven that the train of Bretherton bubbles not only brings the adjacent aqueous streams closer together but also imposes a recirculatory flow in the transverse direction thereby intensifying the micromixing process. To the best of our knowledge, the thermal Bretherton bubble undergoing an elongation-breakup cycle and its effect on micromixing have not been studied yet, and compared to similar micromixing strategies [16,29,30], the unique feature of the present micromixing strategy lies in the fact that the need for a separation unit/bubble collector is eliminated.

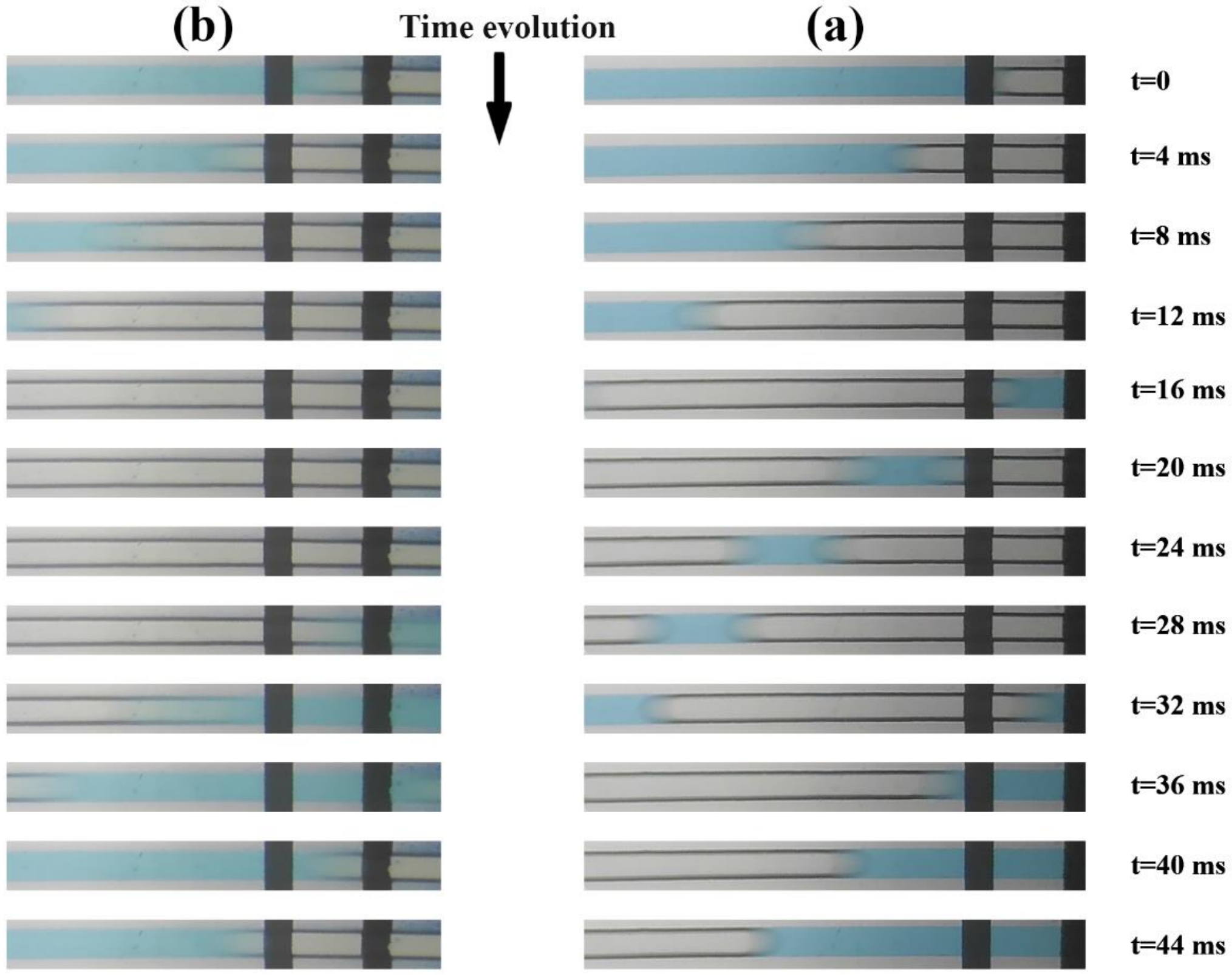

**Fig. 6.** Slug flow pattern in the bubble emergence mixing regime for the flow rate of 20 μL/min: (a) 5 wt% glycerol solution; and (b) 15 wt% glycerol solution.

Nonetheless, whether heating aqueous solutions produces the millisecond-long lifespan elongated bubble (4 and 10 μL/min) or slug flow (20 μL/min), the flow pattern ultimately turns into the bubbly flow at the end of the microchannel, as depicted in Fig. 7. Thanks to the edge detection filter, it is obvious that the satellite bubbles become more disordered by increasing the flow rates. Therefore, the BEMR in the present study can be subcategorized into two sub-regimes: stationary elongation and subsequent pinch-off (for the flow rates of 4 and 10 μL/min) and slug flow pattern with delayed pinch-off (for the flow rate of 20 μL/min).

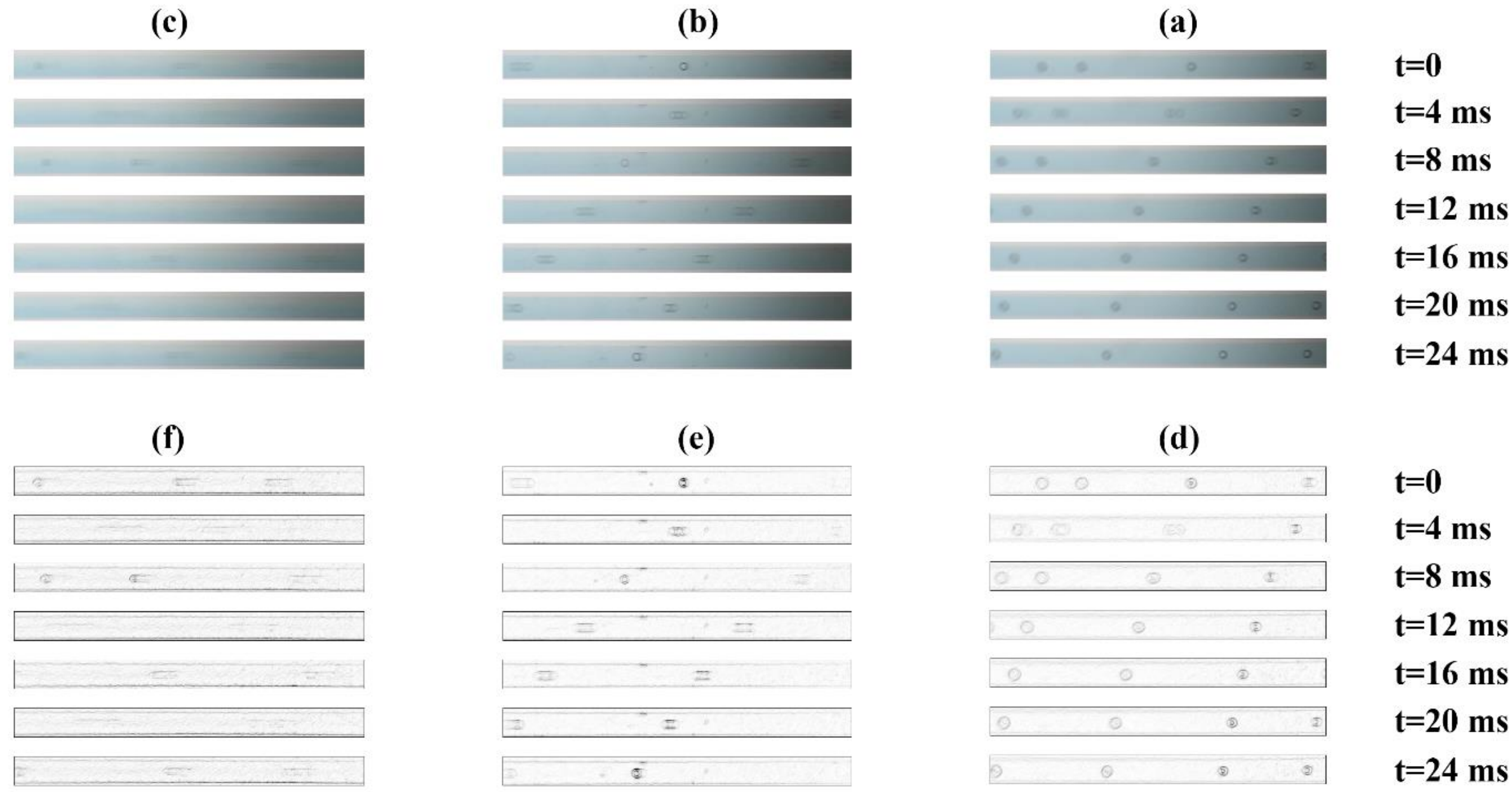

**Fig. 7.** Satellite bubbles at the end of the microchannel for 5 wt% glycerol solution at overall flow rates of: (a) 4 μL/min; (b) 10 μL/min; and (c) 20 μL/min. (d–f) show the edge-detected versions of (a–c), respectively.

According to Fig. 7, it can be deduced that even for the flow rate of 20 μL/min the pinching-off to satellite spherical bubbles occur although the location of it cannot be tracked as easily as the two lower flow rates. Hence, to gain a more precise understanding of the consequences of elongation and pinch-off on the micromixing process, our attention is more focused on the flow rates of 4 and 10 μL/min to easily capture the consequences of the bubble emergence, elongation, and pinch-off cycle on micromixing process. Furthermore, by reviewing the overall picture of these two sub-regimes (shown in Fig. 4 and Fig. 6), one can conclude that the micromixing process based on the BEMR is a function of two effects: (1) the bubble formation, elongation, and its subsequent breakup. (2) a pseudo-stirring role of flowing spherical bubbles toward downstream of the microchannel. While the latter exhibits a path-dependent, progressive mechanism, the former can be regarded as an instantaneous one that exponentially augments mixing by disturbing the otherwise highly ordered liquid-liquid interfaces. Thus, it can be advantageous to classify the micromixing process according to the two mechanisms into two states. Further discussions are presented in the Sections 3.5 and 3.6.

**Table 3**

Measured critical voltage ($CV$), lengths ($L$), growth rates ($\dot{L}$), and nucleation period ($\tau$) of elongated bubbles in two overall flow rates of 4 and 10 μL/min for water and various glycerol-water solutions.

| | | Water | 5 wt% GS | 10 wt% GS | 15 wt% GS | 20 wt% GS |
|---|---|---|---|---|---|---|
| $CV\ (V)$ | $4\ \mu L/min$ | 4.3 | 5.6 | 5.7 | 5.8 | 6.5 |
| | $10\ \mu L/min$ | 4.5 | 5.8 | 5.9 | 6.4 | 7.6 |
| $L\ (\mu m)$ | $4\ \mu L/min$ | $352 \pm 10$ | $414 \pm 13$ | $432 \pm 10$ | $456 \pm 10$ | $470 \pm 12$ |
| | $10\ \mu L/min$ | $465 \pm 14$ | $715 \pm 11$ | $740 \pm 16$ | $768 \pm 13$ | $790 \pm 15$ |
| $\dot{L}\ (mm/s)$ | $4\ \mu L/min$ | $32 \pm 1.6$ | $36 \pm 2$ | $36 \pm 2$ | $36 \pm 3$ | $37 \pm 2$ |
| | $10\ \mu L/min$ | $40 \pm 2$ | $70 \pm 4$ | $73 \pm 5$ | $75 \pm 5$ | $95 \pm 6$ |
| $\tau\ (ms)$ | $4\ \mu L/min$ | $32 \pm 4$ | $36 \pm 4$ | $36 \pm 4$ | $36 \pm 4$ | $20 \pm 4$ |
| | $10\ \mu L/min$ | $16 \pm 4$ | $20 \pm 4$ | $20 \pm 4$ | $16 \pm 4$ | $16 \pm 4$ |

### 3.3 Bubble nucleation

A local increase in the temperature of the substrate suggests a robust affinity for coercing dissolved air within the aqueous streams as well as trapped air in cavities into the otherwise diffusion-constrained flows. As it was discussed by Pereiro et al. [25], flowing liquids through the microfluidic devices usually have a considerable content of dissolved air because of various sources, e.g., routine pumping devices and the high level of permeability of polymers used in microchannels. Volk et al. [26] showed that even when degassed water is used in microfluidic studies, due to the air-permeable pathways within the PDMS walls, it quickly regains its air-saturated condition. From a chemical equilibrium thermodynamic perspective, Henry's law states that the fugacity of dissolved air in a liquid phase, $\hat{f}^{L}_{air}$, can be approximated by $k_H x_{air}$ where $k_H$ is Henry's law constant and $x_{air}$ denotes the molar fraction of air in the liquid phase. $k_H$ is a decreasing function of temperature, meaning that an increase in the temperature of the liquid would compel the dissolved air to migrate from the liquid and thus constituting a nascent gaseous phase. Subsequently, if the reverse condition happens (i.e., a temperature decrease) the liquid has the potential to accumulate more air within itself.

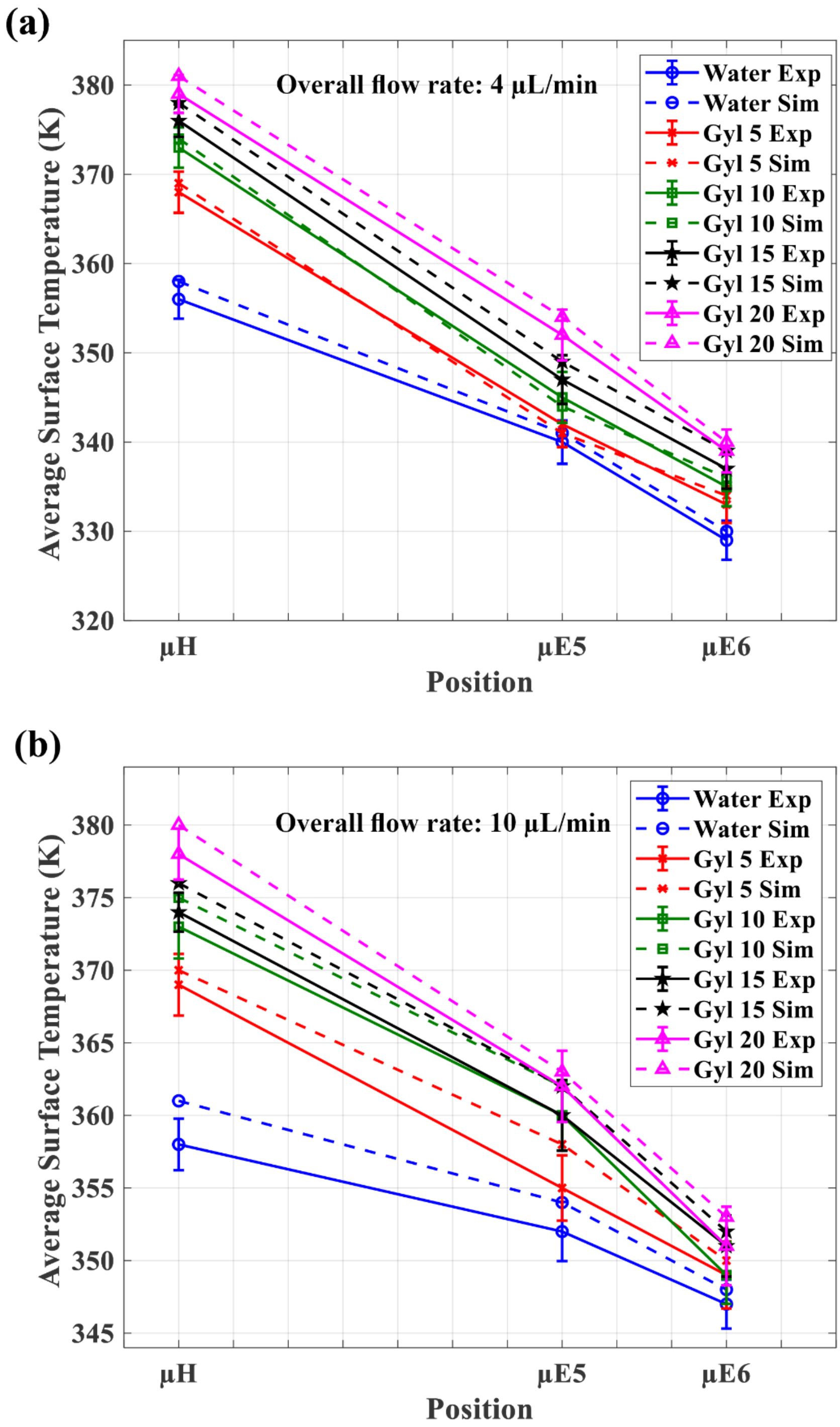

**Fig. 8.** Average surface temperature at the critical voltage for the micro heater (µH) and two micro electrodes (µE5 and µE6) for various aqueous solutions: (a) overall flow rate of 4 µL/min; and (b) overall flow rate of 10 µL/min. Experimental results (Exp) are shown in solid lines while the dashed lines represent simulation results (Sim).

As demonstrated in Fig. 1 (c), the surface of the sputtered platinum µEs contains numerous nano- and micro-scale cavities. There is a general consensus that these cavities serve as active sites for bubble nucleation [25]. Additionally, as the flowing aqueous solutions regularly experience an air-supersaturated condition—due to the high level of air saturation at the inlet and the pressure drop along the conduit—they become highly susceptible to bubble nucleation when the microheater surface temperature surpasses a critical value corresponding to its critical voltage. This process is known as heterogeneous bubble nucleation [25].

In Fig. 8, the average surface temperature at the critical voltage for the microheater and two downstream micro electrodes (μE5 and μE6) for various aqueous solutions at two overall flow rates of 4 and 10 μL/min are presented. The thermometry is also supplemented by a two-dimensional finite element numerical simulation using the COMSOL Multiphysics 5.6 software in order to extend the thermometry results for estimating the temperature around the elongated bubbles. In the simulations, the vanishing liquid film thickness, $\varepsilon$, between the elongated bubble and the wall (as shown in Fig. 2) is chosen in the range of 3-5 μm to be as close as possible to the available theoretical estimations [19,37,39].

According to Fig. 8, the experimental results show that the microheater surface temperature for the occurrence of the BEMR, when side-by-side dyed and clear water flow, can be approximated in the range of $357-360\ K$ which is in acceptable accordance with the results of a microchannel flow boiling investigation performed by Steinke and Kandlikar [45] for undegassed water. For water as the working fluid, due to its high capability of dissolving air (compared to glycerol added aqueous solutions), a rise in the microheater surface temperature to the denoted level ensures heterogeneous nucleation. However, the case is different for aqueous solutions formed by adding glycerol to water. In this regard, as shown in Fig. 8 (a) and (b), bubble nucleation occurs when the microheater wall temperature rises in the range of $375-380\ K$, for a 20 wt% glycerol solution, representing a considerable 20 K increase compared to water.

Although heterogeneous bubble nucleation in microscale cooling applications can be envisioned as one of the primary regimes in boiling heat transfer, it has not received widespread attention more specifically when it comes to low flow rates of the coolant [45–48]. This is partly due to the substantially higher heat fluxes demanded for cooling applications, and partly to the insulation of heater surface [47,48] and, in this regard, a degassing stage is typically carried out before the heat transfer-oriented investigations [49]. However, as demonstrated in this study, thermally induced bubble nucleation can lead to instantaneous, enhanced micromixing and, consequently, improved microscale mass transfer. To this end, investigating the interplays between governing forces acting during the BEMR seems advantageous for recognizing the micromixing process.

### 3.3.1 Brief discussion on bubble composition

It is also worth noting that we also observed an initially anomalous scenario (we refer to it as the evaporation-dominated regime) when the MF chip was used for the first time. The snapshots shown in Fig. 9 belong to the experiment performed for overall water flow rate of 4 μL/min when the MF chip was locally heated up for the first time in which we did not observed any bubble nucleation until the critical voltage surpassed $8.1\ V$ (approximately equivalent to the microheater surface temperature of $383\ K$). In performing the experiments of this paper, we remanufactured the MF chip three times and this scenario, which is atypical at first, occurred when the MF chip was used for the first time. However, it did not recur in the later tests of each MF chip and the critical voltage of bubble nucleation fell significantly below 8.1 V for different liquids and also in different flow rates. According to Fig. 5, the nearest critical voltage for typical experiments in this study is $7.8\ V$ for 20 wt% of glycerol solution when it follows with overall flow rate of 20 μL/min. We attribute the need for higher microheater surface temperature for glycerol solutions to the smaller capacity for dissolving air as discussed in the last paragraphs of the previous section.

As mentioned in the introduction, PDMS is intrinsically hydrophobic. Although plasma bonding treatment makes it nearly hydrophilic, the hydrophobic recovery of this treatment, typically occurring within about one day [50,51], tends to entail the wettability of surfaces in the Cassie-Baxter state for trapped air pockets [25,28]. Hence, we performed the basic experiments after conducting cold tests within the first three days after manufacturing the MF chip to ensure the repeatability of the results. Consequently, due to the three following reasons, we believe that the generated bubbles in our work (those shown in Fig. 4, Fig. 6 and Fig. 7 as well as upcoming discussion and figures excluding those in Fig. 9) are mostly occupied by air while the generated bubbles in the evaporation-dominated regime contain a higher amount of vapor:

(1) For the evaporation-dominated regime (the initially anomalous scenario), the microheater surface temperature must reach above $380\ K$ in order for nucleation to occur which is consistent with the supersaturation temperature difference that typically occurs in micro-scale flow boiling heat transfer [45–49,52]. However, in all experiments performed after the first three days of MF chip manufacturing, nucleation occurs in microheater surface temperatures considerably below that range, particularly for water (see Fig. 8).

(2) Generated bubbles in Fig. 9 are vapor bubbles as the evaporation process is obvious, more specifically in snapshots 24-40; a phenomenon that is almost absent in the overall body of our study.

(3) The timescale of bubble nucleation, growth, and subsequent pinch-off throughout the rest of our study is on the order of milliseconds, whereas elongation and evaporation last for over one second in the evaporation-dominated regime, as shown in Fig. 9.

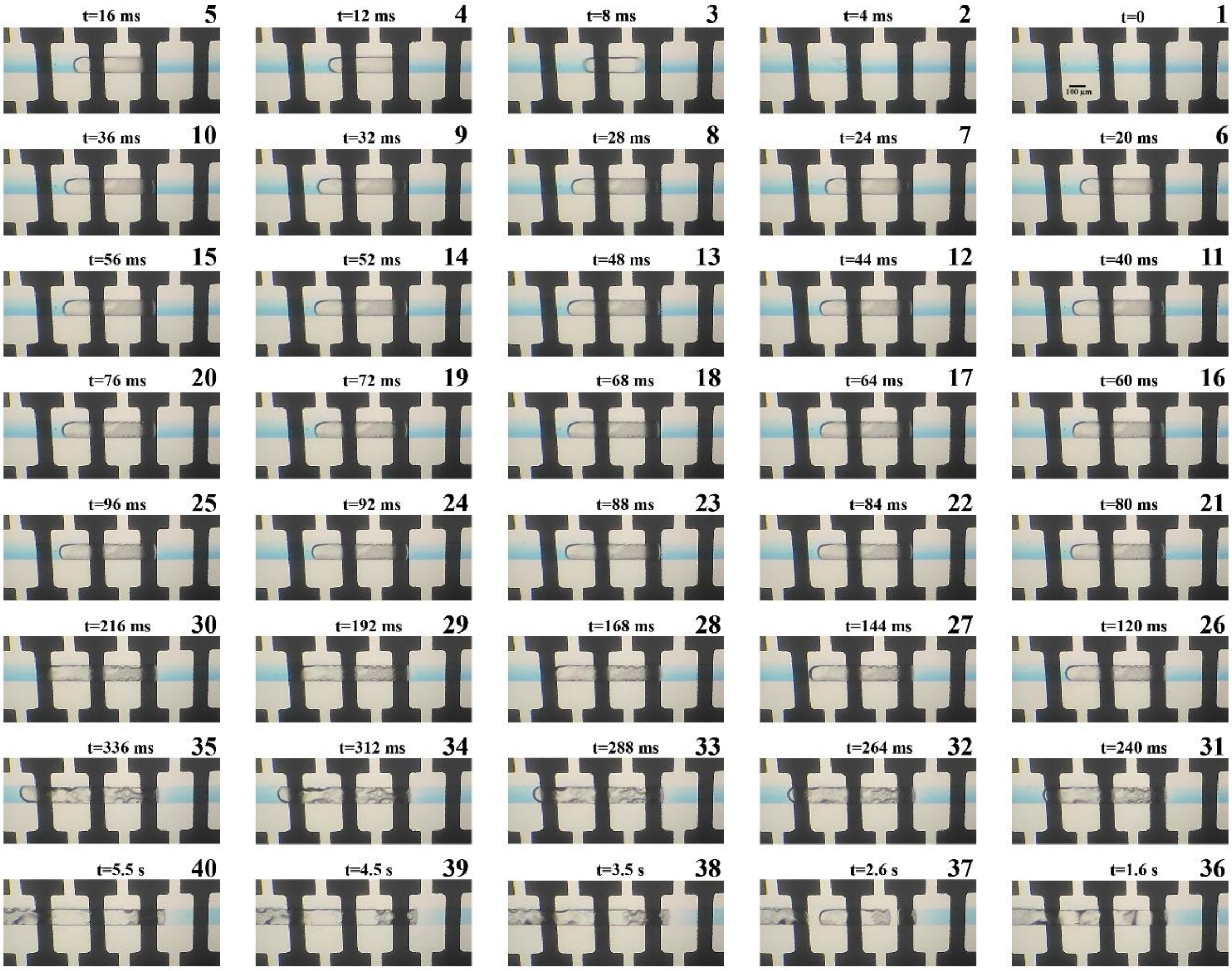

**Fig. 9.** Snapshots of bubble generation for the evaporation-dominated regime (the initially anomalous scenario) when the microfluidic chip is used for the first time at the critical voltage of 8.1 V.

## 3.4 Emergence, elongation, and pinch-off

Upon the nucleation of a bubble on the microheater surface, it grows and nearly fills the cross-section of the microchannel while also expanding axially, provided that the necessary thermal conditions are consistently maintained. During early stages of bubble growth and filling the cross-section of the channel, one may draw a connection to the T-junction microfluidic bubble generation [42], more specifically its squeezing regime, by considering the forming thermal bubble as an injecting gaseous phase into the background flow. There is a consensus that the competition of two forces governs the properties of generated bubbles via T-junctions: (1) The enormous pressure drop of the squeezing film between the forming gas bubble and the wall, and therefore, its associated resistive force. (2) The capillary pressure gradient force; a force related to the existence of curvature gradient along the forming bubble. Although supposing this similarity for the early stages of bubble growth and filling the cross-section of the channel seems plausible, this analogy intrinsically fails to mimic further stages of the thermal bubble.

Since the thermal bubbles in the present micromixing scheme are essentially shaped as a result of the thermally excited air within the system, disappearance of sufficient temperature rise around the elongated thermal bubbles would conversely influence them to remigrate to the surrounding flow whose temperature is now significantly lower than the microheater surface. To quantitatively view the described scenario, the temperature fields for two flow rates of 4 and 10 μL/min of water are depicted in Fig. 10 (a) and (b), as obtained from the numerical simulation. Due to its low thermal conductivity, air acts as an insulator, with the nearest squeezing films serving as the only means for convecting the generated heat from the microheater surface.

Meanwhile, as the bubble becomes more elongated, its advancing front experiences lower temperatures causing the encapsulated pressurized air within the bubble to return to the surrounding fluid. This stage culminates in the disintegration of the bubble into satellite daughter bubbles and microjetting, both of which contribute to intensifying the micromixing process through their stirring effect. A similar scenario is also expected to occur for the higher flow rate (i.e., 20 μL/min) with some differences; the bubble becomes longer, moves downstream, and due to higher convective flow, its breakup is delayed to a more distant position. Compared to isothermal bubble-assisted strategies [16,29,30], it is now clear why there is no need for an extra bubble separation unit.

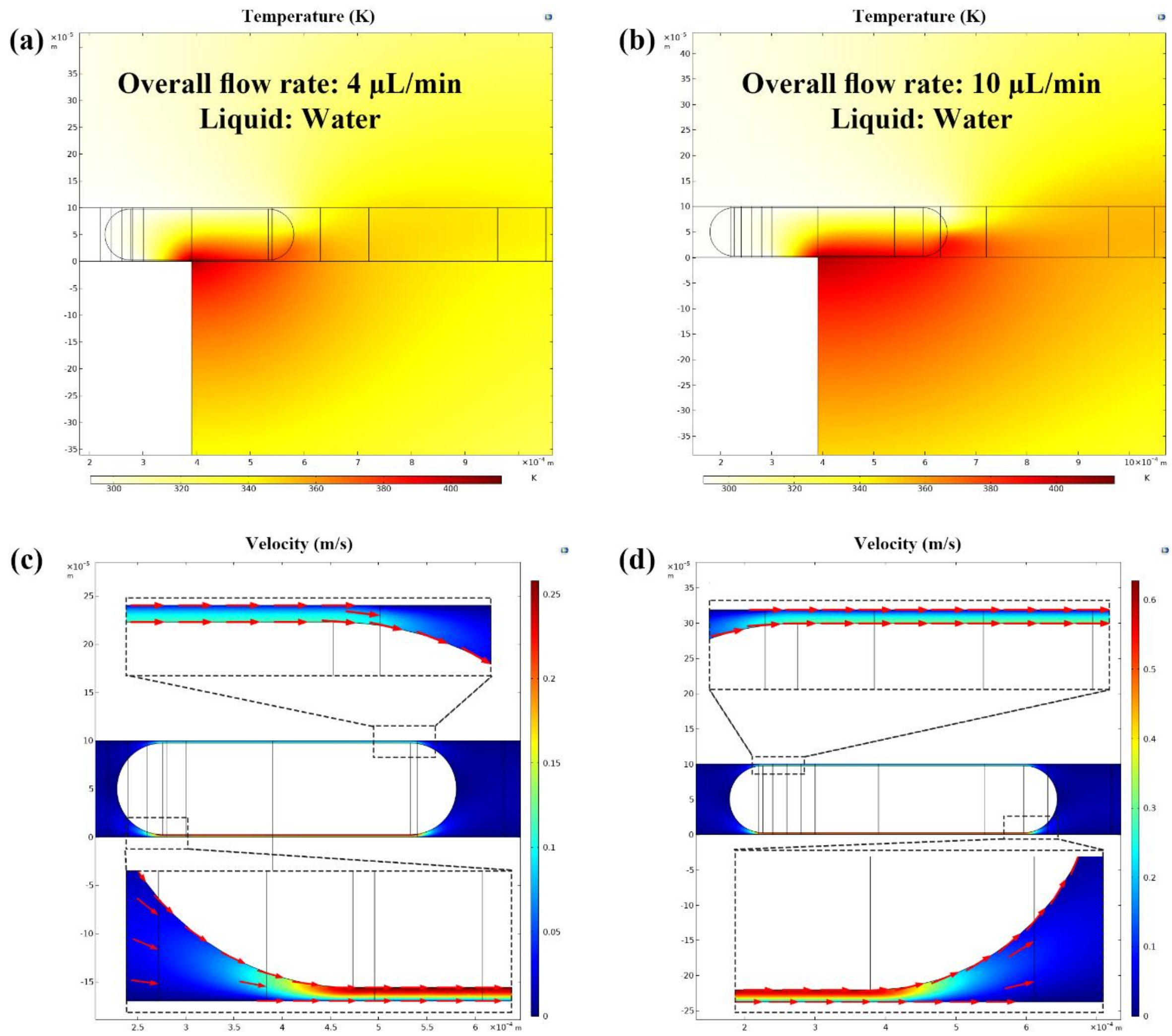

**Fig. 10.** The liquid temperature fields around the elongated bubbles in two different overall flow rates: (a) 4 μL/min; and (b) 10 μL/min. The corresponding velocity fields are also shown in (c) and (d), respectively.

The time evolving micrographs of the formation, elongation and subsequent pinch-off for the case of 5 wt% glycerol solution and the overall flow rate of 4 μL/min are shown in Fig. 11 (a) in color and Fig. 11 (b) in a grayscale, background-removed representation (see also the supplementary video where water is the working fluid). Compared to Fig. 4, due to the lower flow rate and higher viscosity, the sequence of phenomena is markedly slowed down, facilitating the interpretation based on the 4 ms resolution. Through image processing and using first- and second-order forward numerical differentiation method (considering the minimum time resolution of 4 ms), the growth rate of the elongating bubbles at the onset of formation is calculated, and the results are presented in Table 3 for the overall flow rates of 4 and 10 μL/min. It should be noted that because finer time resolution is required, the corresponding results for the overall flow rate of 20 μL/min are not calculated. According to Fig. 11, until 16 ms after the emergence of the bubble, it experiences the elongation mode where it reaches an approximate length of 414 μm.

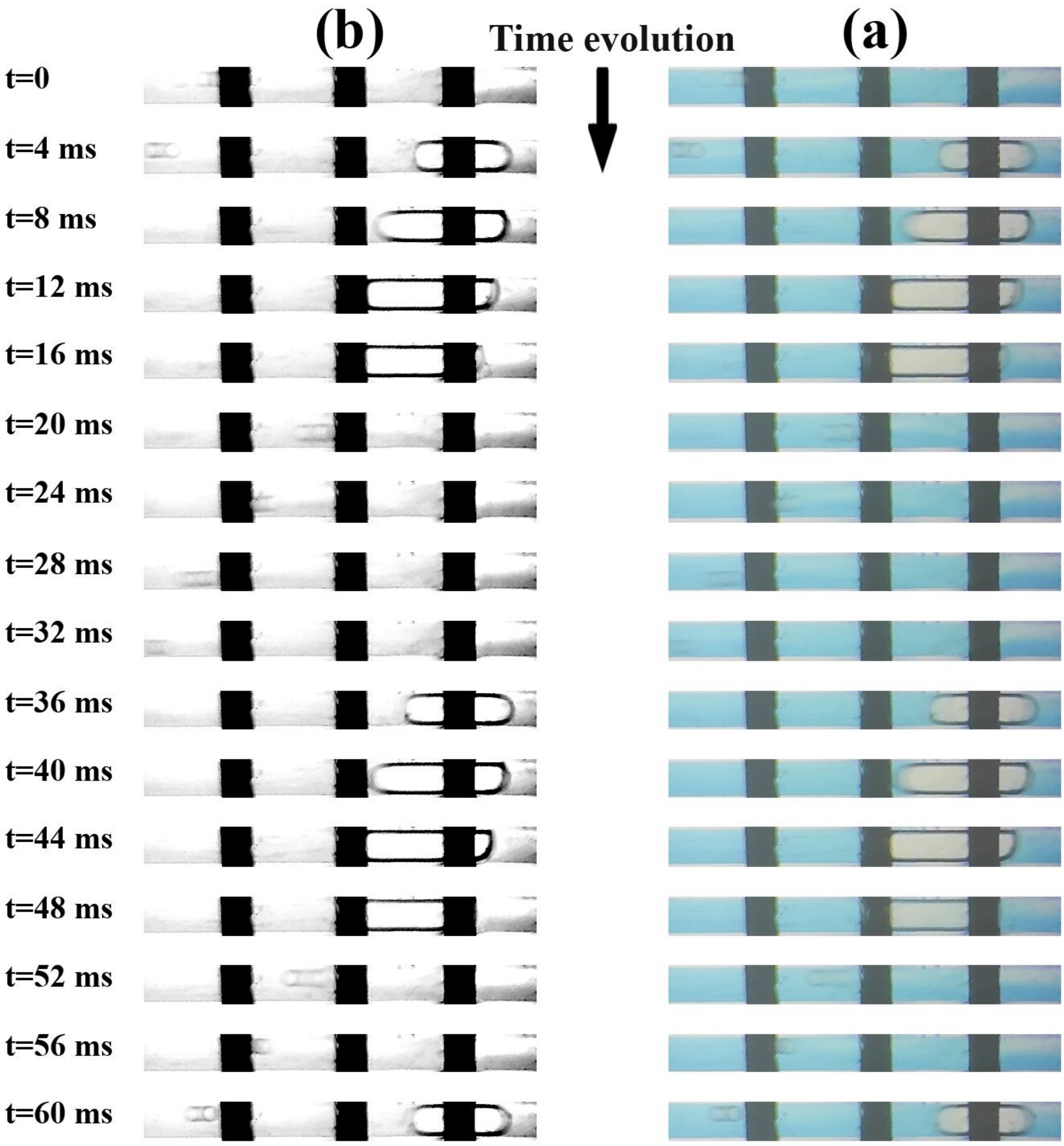

**Fig. 11.** Formation and subsequent pinch-off of the elongated bubble, shown in both color (a) and a grayscale, background-removed representation (b). The flowing solutions consist of dyed and clear 5 wt% glycerol in water, with an overall flow rate of 4 μL/min.

As it was discussed, the emergence and elongation of the thermal bubbles significantly enhances the micromixing process, more specifically ahead of the bubble. After the breakup and disintegrating into satellite bubbles, the spherical bubbles move downstream while playing a stirring role to even more increase the mixing index. In this regard, the MIs ahead of the microheater (see Fig. 3 and Fig. 5) are mainly influenced by the interactions between four forces/stresses:

- The momentum force of the growing bubble (primarily due to $\dot{L}$ in Table 3).
- The viscous stresses in the squeezing film (see Fig. 10 (c) and (d)).
- The thermal Marangoni tensile stresses across and along the elongating bubbles (as depicted in Fig. 12).
- The capillary force, specifically in the nose and the rear of the elongated bubble where the curvature gradient is significant.

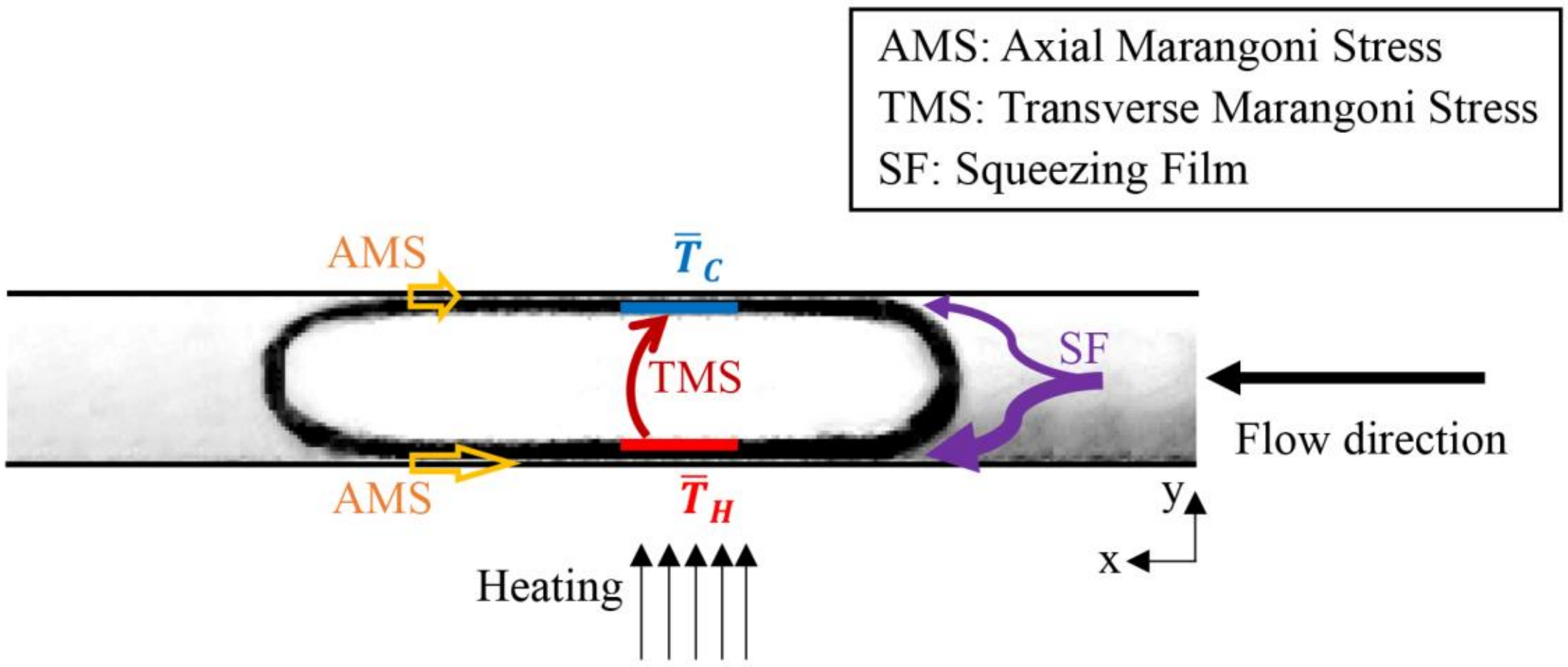

**Fig. 12.** Schematic of the competition between the squeezing film and the Marangoni stresses during the bubble emerging mixing regime.

Furthermore, the slight enhancements in the MIs at the end of the microchannel can be explained through the following two arguments:

- The convective effect of streaming satellite bubbles which tends to disturb the mixing barriers.
- The capillary force around streaming satellite bubbles which tends to preserve the spherical shape.

The scaling arguments as well as quantitative relationships between the influencing forces for the MIs ahead of the microheater are represented in Section 3.5, while in Section 3.6 the governing non-dimensional groups in the MIs at the end of the microchannel are discussed.

## 3.5 Micromixing mechanism during bubble elongation pinch-off

### 3.5.1 Scaling argument for the Marangoni number

The elongation of bubbles in the media with the temperature gradient not only impacts the micromixing process through an oscillatory momentum-driven flow of elongated bubbles, but also through a net tensile thermal Marangoni stresses pulling the air-liquid interfaces from the warmer sections toward the colder ones, as shown in Fig. 12; the transverse Marangoni stress is due to the temperature difference $\bar{T}_H - \bar{T}_C$ occurring over a length scale of $\mathcal{O}(h)$ whereas the axial Marangoni stress arises from a lower temperature difference between the nose and the rear of the elongated bubble (compared to $\bar{T}_H - \bar{T}_C$ as shown in Fig. 10) over a length scale of $\mathcal{O}(L)$ ($\mathcal{O}(L) > \mathcal{O}(h)$). The later Marangoni stress should also confront the advancing squeezing film whose mean velocity is $\bar{U}_{SF}$ which can be scaled as $\mathcal{O}(Q/h\varepsilon)$ ($Q$ and $\varepsilon$ are the flow rate and the film thickness, respectively). The temperature difference, $\bar{T}_H - \bar{T}_C$, for various aqueous solutions at two flow rates of 4 and 10 μL/min are presented in Fig. 13. As expected, the estimated values for $\bar{T}_H - \bar{T}_C$ are in the order of the temperature difference between the microheater temperature (see Fig. 8) and the inlet temperature of aqueous solutions, i.e., 20 °C.

Based on the tangential force balance at the liquid-bubble interface, one can conclude that the difference between the shear stresses at the interface should be balanced by a tensile stress, namely the tangential gradient of surface tension (thermal-Marangoni stress) [53],

$$\tau_{liq} - \tau_{bub} = \nabla_s \sigma. \tag{3}$$

In Eq. (3) $\tau$ is used for shear stresses, whereas $\nabla_s$ represents the tangential gradient operator across/along the interface, respectively. The subscripts *liq* and *bub* refer to liquid and bubble. According to the left-hand side of Eq. (3), one can consider the $\tau_{liq} = \mu_{liq}\dot{\gamma}$ as the scale of the viscous shear stress of the squeezing film where $\dot{\gamma}$ denotes the rate of shear strain. Considering the higher temperature difference over a lower distance, we are more interested in the scaling comparison between the transverse Marangoni stress and the viscous shear stress of squeezing films. To build this scaling argument more quantitatively, the Marangoni number ($Ma$) which stems from a scaling view of Eq. (3), is defined as follows:

$$Ma = \frac{\text{The transverse gradient of surface tension}}{\text{The shear stress of the squeezing film}}.$$

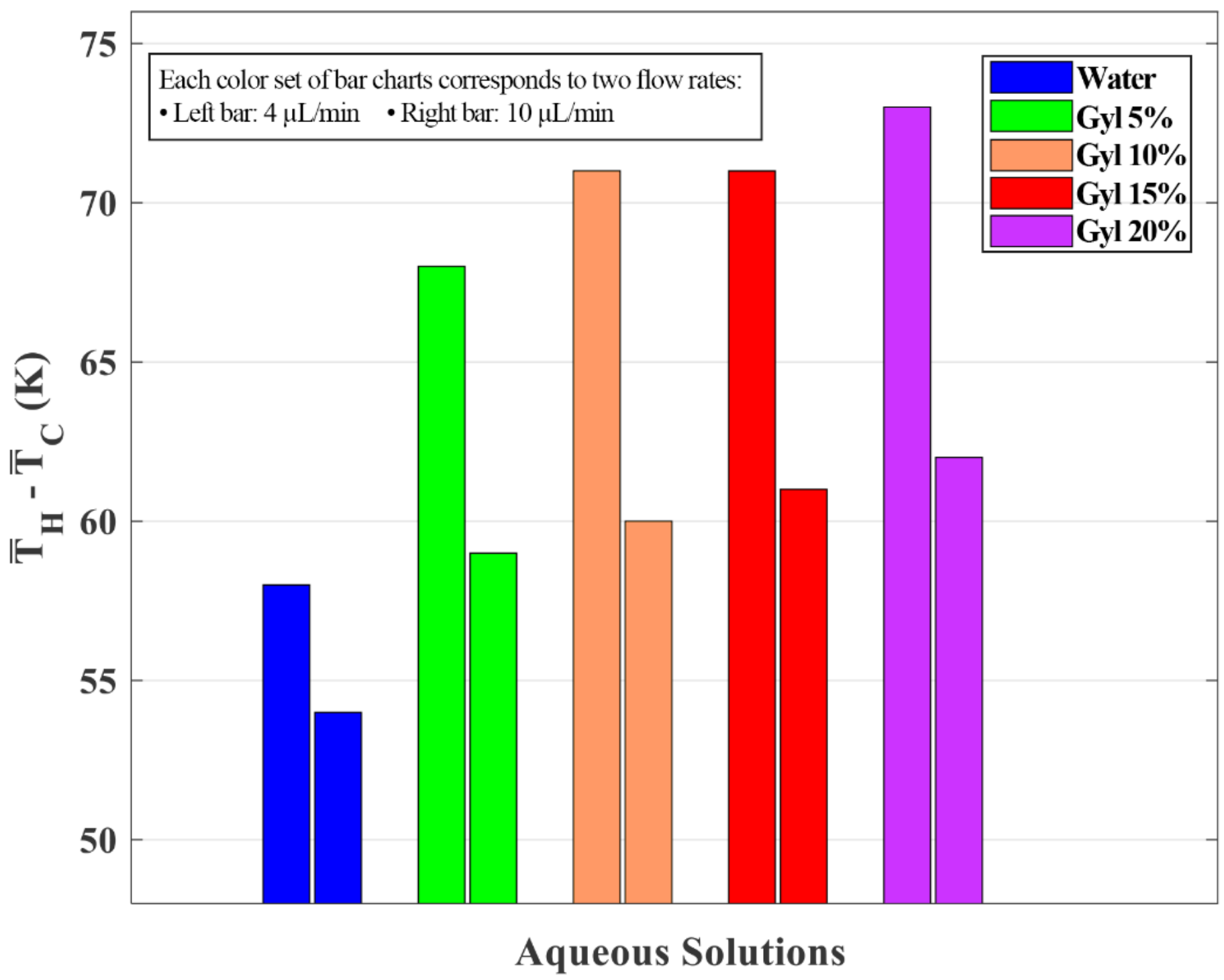

**Fig. 13.** Temperature difference, $\overline{\boldsymbol{T}}_{\boldsymbol{H}}$ and $\overline{\boldsymbol{T}}_{\boldsymbol{C}}$, around emerged elongated bubbles resulted from simulations of various solutions at two overall flow rates of 4 and 10 μL/min.

By representing the transverse gradient of surface tension as $\sigma_T(dT/ds)$ ($\sigma_T$ is the temperature-coefficient of surface tension, $T$ is the temperature, and $s$ denotes the transverse direction pointing from $\overline{T}_H$ toward $\overline{T}_C$, as shown in Fig. 12), one can represent the $Ma$ as:

$$Ma = \frac{\sigma_T\left(\frac{dT}{ds}\right)}{\mu_{liq}\dot{\gamma}} \approx -\frac{\sigma_T\left(\frac{\overline{T}_H - \overline{T}_C}{h}\right)}{\mu_{liq}\left(\frac{\overline{U}_{SF}}{\varepsilon}\right)}. \tag{4}$$

By utilizing the well-known universal scaling argument for the squeezing film thickness, $\varepsilon$, according to the Bretherton coating flows [19,37], one can consider:

$$\frac{\varepsilon}{h} \propto Ca^{2/3}. \tag{5}$$

Now, we should seek an explicit argument for $\varepsilon$ to insert within Eq. (4). In this regard, it seems beneficial to consider $Ca$ based on the squeezing film and place the mean velocity of the squeezing film, $\overline{U}_{SF}$, into it:

$$\varepsilon^5 \propto h\left(\frac{\mu_{liq}Q}{\sigma}\right)^2. \tag{6}$$

Therefore, by combining Eq. (4) and Eq. (6) and recalling that $\mathcal{O}(\bar{U}_{SF}) = \mathcal{O}(Q/h\varepsilon)$, the final explicit version of $Ma$ is represented in Eq. (7). It is apparent that through the specific, physics-inspired definition of $Ma$ in the present study, the importance of $Ca$ is also considered for studying $MI$ ahead of the microheater.

$$Ma = -\frac{\sigma_T\left(\frac{\bar{T}_H - \bar{T}_C}{h}\right)}{\left(\frac{\sigma^4 \mu Q}{h^7}\right)^{1/5}}. \quad (7)$$

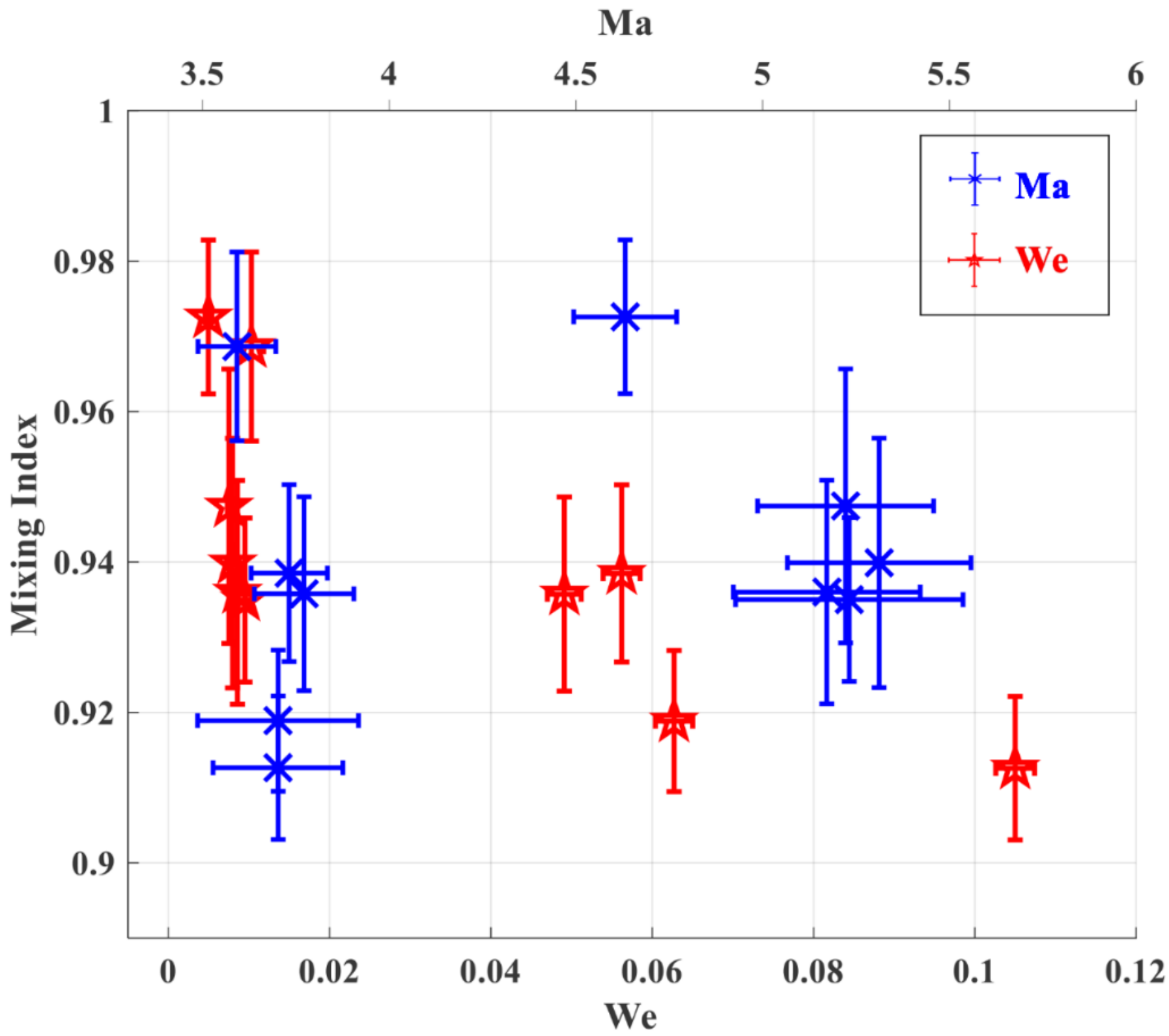

**Fig. 14.** The mixing index ahead of the microheater versus the Weber number (in red) and the Marangoni (in blue).

Accordingly, the mixing indices ahead of the microheater versus the Marangoni numbers of the corresponding data (Eq. (7)) are plotted in Fig. 14. One interesting feature of the represented scaling argument in this section is that the corresponding values of $Ma$ lie in the scale of unity demonstrating that both the numerator and the denominator of $Ma$ are of the same order. It should also be noted that in the $MI$ data for flow rates of 4 and 10 μL/min are considered, as previously discussed in Section 3.4. As it is obvious in Fig. 14, $Ma$ is not the sole independent non-dimensional variable governing the variations of the $MI$. Therefore, in the next step, we aim to define the Weber number ($We$) as another important non-dimensional group in micromixing while the final concluding paragraphs of the next section discuss the corresponding variation of $MI$ with both $We$ and $Ma$.

### 3.5.2 Effect of the Weber number

The existence of the elongated bubble creates an ephemeral partial blockage near the microheater which enhances the micromixing process in two ways: Firstly, it induces recirculation to the otherwise axially aligned liquid-liquid interface between the two streams. Secondly, it enforces the squeezing films to undergo a pseudo one-stage separation and recombination process. In this regard, the final length of the elongated bubbles before breakup ($L$ in Table 3) along with their corresponding growth rate (i.e., $\dot{L}$) can be considered two influential parameters in the micromixing process. According to the Rayleigh-Plesset equation [53] and also inspired by the way Hoeve et al. [54] defined the Weber number ($We$), the relative effect of the momentum force to the surface tension can be demonstrated as:

$$We = \frac{\rho\left(\dot{L}\right)^2 L}{\sigma}. \tag{8}$$

The mixing indices ahead of the microheater versus the Weber numbers of the corresponding data are also plotted in Fig. 14 showing an approximate decreasing relationship between $MI$ and $We$. Accordingly, the BEMR can be classified into two sub-regimes, namely $We < 0.02$ and $We > 0.04$. As is also evident in Fig. 14, for the former sub-regime, micromixing is more influential in $Ma$ where a decrease in $MI$ from 97 % to 92 % occurs for variation of $Ma$ from 3.5 to 4; the low values of $We$ can be interpreted as the domination of surface tension forces over the inertia. The domination of surface tension forces influences the micromixing process in two opposing ways: (1) It acts as a retarding force against the inertia providing a wider opportunity for the Marangoni stresses to impose a vortical flow to the surrounding streams (enhancing participation). (2) As is evident from the definition of $Ma$ (see Eq. (7)), surface tension, which tends to minimize the interfacial area, defies the Marangoni stresses through strengthening the shear stresses (decreasing participation).

On the other hand, when $We$ exceeds 0.05, the micromixing mostly becomes dependent on $We$, leading to a decrease in $MI$ from 94 % to 91 % as $We$ increases from 0.05 to 0.11; as $We$ increases, the surface becomes flatter, indicating that inertial effects reduce the sensitivity of $MI$ to variations in $Ma$. In this sub-regime, initially, inertia alters the balance of participating forces and enhances the micromixing. This condition can be attributed to the fact that inertia diminishes the impact of the shear stresses. However, after this transitional region, inertia predominantly reduces micromixing. As mentioned earlier, since the time resolution as well as the field of view used for visualization of experiments with the overall flow rates of 4 and 10 µL/min were insufficient to capture the required $We$ and $Ma$ parameters at overall flow rates of 20 µL/min or higher, we could not include more data and explore the functional relationship for higher values of $We$ and $Ma$. Furthermore, we believe the presented functional relationship could be refined by conducting more tests to cover a broader range of $We$ and $Ma$ variations and better capture the transition region between the two sub-regimes.

### 3.6 Micromixing mechanism at the end of the microchannel

As shown in Fig. 5, when the BEMR occurs, $MI$ in all aqueous solutions increases at the end of the microchannel compared to its counterpart ahead of the microheater; for example, in the case of a 10 wt% glycerol solution, increases of 3.5 %, 2.0 %, and 2.2 % are observed for the overall flow rates of 4, 10, and 20 µL/min, respectively. In addition to the fact that streams flow approximately 14 mm further, the presence of moving spherical satellite bubbles can be considered the primary reason for this mixing enhancement. Therefore, to analyze this behavior through the non-dimensional groups, the Péclet number (Pé)—the ratio of the convective mass transfer to the diffusive mass transfer—is investigated. In order to involve the role of flowing bubbles, the Péclet number is defined as follows, inspired by the work of Shim et al. [55] and Groß and Pelz [56]:

$$P\acute{e} = \frac{\dot{\gamma} d^2}{D_{AB}}, \tag{9}$$

where $\dot{\gamma}$ is scaled by $\dot{\gamma} \approx U/h$. The average diameter of the following bubbles is denoted by $d$, calculated based on processing the captured images from the end of the microchannel. In addition to $P\acute{e}$, Ca is also considered another influential non-dimensional group. In, $MI$ at the end of the microchannel is plotted against the corresponding $P\acute{e}$ and $Ca$ values. Accordingly, Fig. 15 implies two distinct physical mechanisms that affect $MI$ further compared to its values ahead of the microheater, almost separately. As also indicated by the definition of $P\acute{e}$, the convective effect augments the micromixing process due to the displacement of spherical satellite bubbles downstream. Furthermore, as shown in Fig. 7, the increasement of $Ca$ which is mostly sensitive to increasement of the flow rate, reflects itself in reduction of bubble diameter while also increasing its translational velocity.

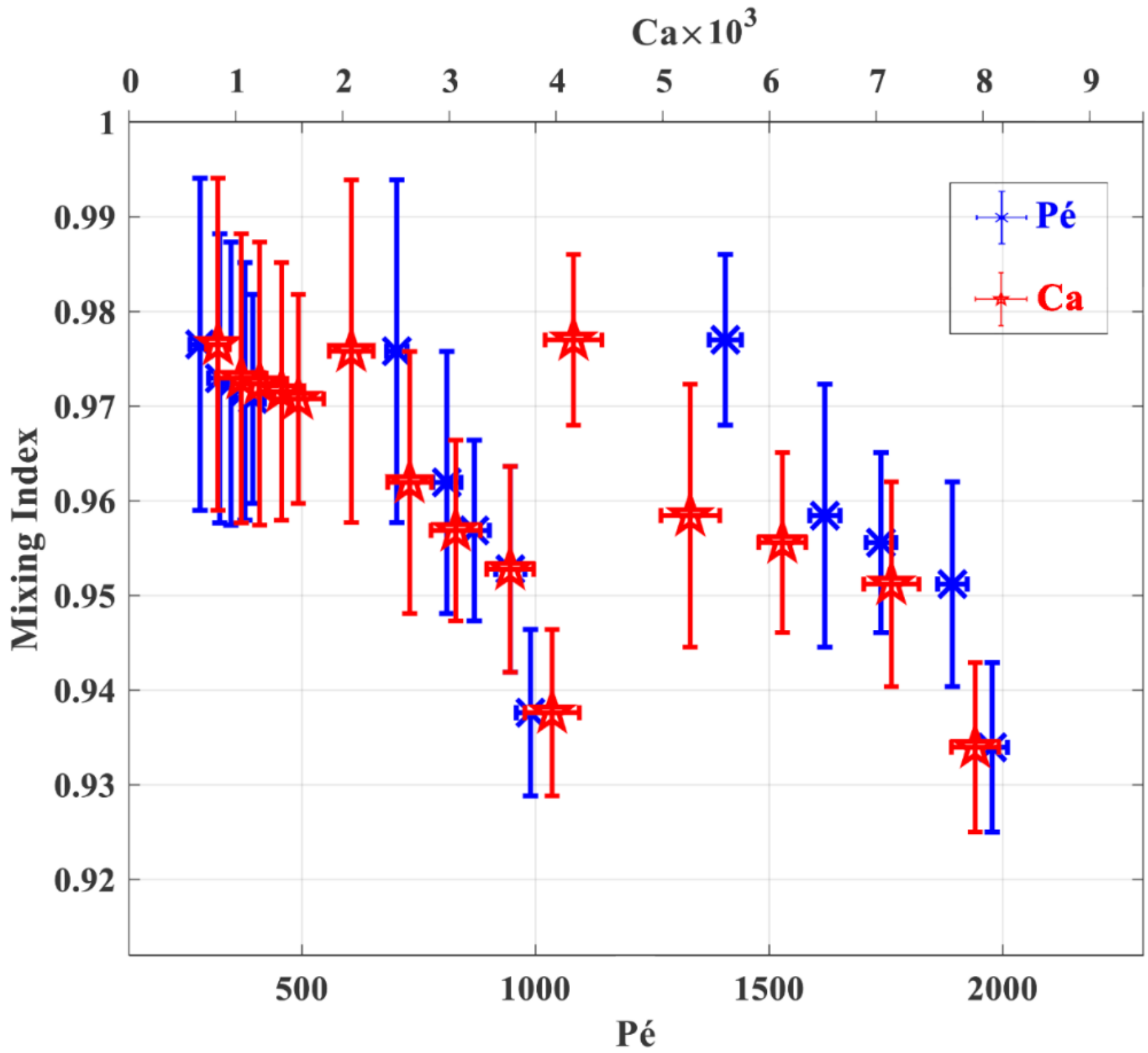

**Fig. 15.** Downstream mixing index versus the Péclet number (in red) and the Capillary number (in red).

## 4 Conclusions

In this paper, a bubble-assisted micromixing strategy was presented based on thermally exciting the intrinsic air within the microchannel crevices and the fact that the flowing streams in microchannels made by PDMS are in air-saturated conditions. Thermal excitation was achieved through the direct contact between the sputtered thin platinum microheater and the co-flowing streams. The investigation involved examining five different aqueous solutions (i.e., from pure water to 20 wt% glycerol solution) as well as three different overall flow rates: 4, 10, and 20 μL/min. It was revealed that before the emergence of bubbles on the microheater surface, the HIMR is responsible for the micromixing enhancement. However, when the microheater temperature exceeds a critical value, a bubble emerges and, in less than 4 ms, fills the microchannel cross-section and then tends to elongate along the microchannel. Furthermore, the BEMR is responsible for almost thorough mixing ahead of the microheater ($MI$ reaches more than 95 % for water, and above 91 % in the worst-case scenario of a 20 wt% glycerol solution flowing at 20 μL/min). Overall, we believe the presented micromixing strategy can be easily adopted for prefabricated MF masks, where it will robustly provide the benefits of segmented flow regimes. The major results are recapped as below:

- Based on the different flow rates, two main hydrodynamics flow patterns were identified, namely the millisecond-long lifespan elongated bubble (at 4 and 10 μL/min) and slug flow (at 20 μL/min). In the former pattern, the elongated bubble does not move downstream with the background flow, and it disintegrates near the microheater into satellite spherical bubbles. In the later pattern, however, slug flow pattern occurs, and the pinch-off and the microjetting process shift further downstream.
- In both flow patterns, the satellite spherical bubbles at the end of the microchannel (see Fig. 7) are responsible for the further enhancement of mixing due to their stirring effect on the flowing liquids.
- For the case of two side-by-side streams of water (dyed and clear), the microheater surface temperature (at the BEMR) reaches around 360 K, where the nucleate bubble emergence occurs. However, this temperature is significantly higher for glycerol-water solutions; for example, it reaches 380 K for a 20 wt% glycerol solution, which is a 20 K increase compared to water.
- The BEMR itself was classified into two sub-regions: ahead of the microheater and at the end of the microchannel. In the first region, through a scaling argument for the bubble-liquid interface, we showed that competition between the transvers Marangoni stress and the viscous shear stress of the squeezing film in the narrow passages between the bubble and the microchannel walls is the key mechanism controlling the local micromixing process. In this regard, after deriving a modified Marangoni number ($Ma$) suitable for the present problem, the effects of the Weber number ($We$) as well as $Ma$ on micromixing were investigated. For $We < 0.02$, it was found that micromixing is more sensitive to $Ma$, while for $We > 0.04$, the effect of $We$ becomes more dominant.
- In the second region of the BEMR, it was found that for the almost entire case studies examined in the present study, both the Capillary number ($Ca$) and the Péclet number ($Pé$) are important to characterize the enhancing effect occurring at the end of microchannel.

# 5 Declaration of generative AI and AI-assisted technologies in the writing process

During the preparation of this work and in a few sentences, the authors used ChatGPT to ensure that the grammatical structures used to express their intentions in the paper are aligned with standard English grammar. After using this tool, the authors reviewed and edited the content as needed and take full responsibility for the content of the published article.

## 6 Acknowledgment

The authors express their sincere gratitude to the students and staff of the “Energy, Water, and Environment Research Center” at the School of Mechanical Engineering, Iran University of Science and Technology, for their valuable support to this work. We also appreciate the technical assistance of Dr. Mahdi Moghimi, Mr. Naeem Jalali and Mr. Adel Mahpour in the microchannel fabrication and setup preparation.